\documentclass[journal]{IEEEtran}
\usepackage{graphicx}
\usepackage{epstopdf}
\usepackage{algorithm}
\usepackage{moreverb}
\usepackage{amsmath}
\usepackage{colortbl}
\usepackage{wrapfig}
\usepackage{multirow}
\usepackage{algorithmic}
\usepackage{tabularx}
\usepackage{cite}
\usepackage{graphicx}
\usepackage{amsfonts}
\usepackage{textcomp}
\usepackage{marvosym}
\usepackage{amsthm}
\usepackage{subfigure}
\usepackage{bm}
\usepackage{amsfonts}
\usepackage{enumerate}
\newtheorem{definition}{Definition}

\newtheorem{theorem}{Theorem}

\usepackage{amsmath}
\usepackage{color}

\begin{document}

\markboth{}
{Shell \MakeLowercase{\textit{et al.}}: Bare Demo of IEEEtran.cls for IEEE Journals}

\title{Two-timescale Resource Allocation for Automated Networks in IIoT}

\author{Yanhua~He,~\IEEEmembership{Member,~IEEE},Yun~Ren,
Zhenyu~Zhou,~\IEEEmembership{Senior Member,~IEEE,}
Shahid Mumtaz,~\IEEEmembership{Senior Member,~IEEE,}
Saba Al-Rubaye,~\IEEEmembership{Senior Member,~IEEE,}
Antonios Tsourdos,~\IEEEmembership{Member,~IEEE,}
and Octavia A. Dobre,~\IEEEmembership{Fellow,~IEEE}.

\thanks{Y. He and Y. Ren are with the State Grid Zhejiang Electric Power Company Ningbo Bureau, Zhejiang, China (E-mail: yanhuahe91927@163.com, ren\_yun@zj.sgcc.com.cn).}
\thanks{Z. Zhou is with the School of Electrical and Electronic Engineering, North China Electric Power University, Beijing 102206, China (E-mail: zhenyu\_zhou@ncepu.edu.cn). Z. Zhou is the corresponding author.}
\thanks{S. Mumtaz is with the Instituto de Telecomunica\c{c}\~{o}es,1049-001 Aveiro, Portugal (E-mail: smumtaz@av.it.pt).}
\thanks{S. Al-Rubaye and A. Tsourdos are with the School of Aerospace, Transport and Manufacturing, Cranfield University, UK (E-mail: s.alrubaye@cranfield.ac.uk, a.tsourdos@cranfield.ac.uk).}
\thanks{O. A. Dobre is with the Faculty of Engineering and Applied Science, Memorial University, Newfoundland, Canada (E-mail: odobre@mun.ca).}
\thanks{Part of this work was presented at 2019 IEEE Global Communications Conference (GLOBECOM), 9-13 December 2019. \emph{(Corresponding author: Zhenyu Zhou.)}}
}

\maketitle
%摘要

\begin{abstract}
The rapid technological advances of cellular technologies will revolutionize network automation in industrial internet of things (IIoT). In this paper, we investigate the two-timescale resource allocation problem in IIoT networks with hybrid energy supply, where temporal variations of energy harvesting (EH), electricity price, channel state, and data arrival exhibit different granularity. The formulated problem consists of energy management at a large timescale, as well as rate control, channel selection, and power allocation at a small timescale. To address this challenge, we develop an online solution to guarantee bounded performance deviation with only causal information. Specifically, Lyapunov optimization is leveraged to transform the long-term stochastic optimization problem into a series of short-term deterministic optimization problems. Then, a low-complexity rate control algorithm is developed based on alternating direction method of multipliers (ADMM), which accelerates the convergence speed via the decomposition-coordination approach. Next, the joint channel selection and power allocation problem is transformed into a one-to-many matching problem, and solved by the proposed price-based matching with quota restriction. Finally, the proposed algorithm is verified through simulations under various system configurations.
\end{abstract}


\IEEEpeerreviewmaketitle

%引言
\section{Introduction}
\label{intro}

\subsection{Background and Motivation}

\IEEEPARstart{A}utomated networks rely on seamless integration of advanced self-optimized techniques to improve efficiency, reliability, and operation economics for industrial internet of things (IIoT) applications \cite{zhou2018robust}. Fifth-generation (5G) cellular technologies provide more resilient network infrastructure for connecting massive IIoT devices. However, carbon dioxide generated by powering cellular infrastructures puts tremendous pressure on the sustainability of 5G-empowered IIoT networks. Faced with the urgent need of green cellular networks, researchers have focused on energy-saving strategies on both data transmission side and energy supply side.

On data transmission side, network sleeping \cite{Liu2016Small} and energy-efficient resource allocation techniques \cite{Orsino2017Time} are widely mentioned, applied, and continuously improved. On energy supply side, harvesting renewable energy such as solar and wind energy is advocated to power base stations (BSs) \cite{Chia2014Energy}. However, renewable energy sources with intermittent and fluctuating characteristics have a large impact on reliable BS operation, which may further affect quality of service (QoS) guarantees. A more feasible approach is to utilize both unreliable renewable energy sources and reliable grid power in a complementary manner \cite{Liu2015Backhaul, Ahmed2013Power}. In this sense, the coexistence of various energy sources further complicates resource allocation in 5G-empowered IIoT networks. There exist several challenges that remain unsolved.

{First, energy resource allocation and communication resource allocation are intertwined with each other, and the joint optimization problem is NP-hard due to the coupling between energy and communication domains. Second, energy resource allocation and communication resource allocation have different granularities. Generally, energy domain information such as energy harvesting (EH) and electricity price changes in a large timescale such as minutes \cite{Guo2013Decentralized}, while communication domain information such as channel state and data arrival changes in a small timescale such as seconds or even milliseconds \cite{Zhang2014Dynamic}. Third, communication resource allocation with long-term constraint involves the coupling among different time slots as well as the coupling between different layers, e.g., rate control in the network layer and power allocation in the physical layer. Existing works on either single-layer performance or short-term deterministic optimization cannot be applied. Last but not least, the large-scale deployment of IIoT devices brings complexity issues. Compared with mobile devices and applications, IIoT devices are usually constrained by limited physical space, energy, communication and computing resources, and IIoT applications have stringent requirements on operation delay and reliability. Therefore, it is important to reduce complexity to cope with numerous implementation constraints and strict operation demands.}

The joint optimization of energy and communication resource allocation in renewable energy powered cellular networks has attracted intensive attentions \cite{Chin2012, Shan2016Energy, Doost2017}. Nevertheless, these researches mainly target at single-timescale resource allocation. There are some works taking different time granularities into consideration. In \cite{Jie2018}, Gong \emph{et al.} studied the timescale difference between energy arrival variation and channel fading, and proposed a low-complexity two-stage joint power allocation and energy management optimization algorithm based on Markov decision process (MDP) and dynamic programming. In \cite{Dantong2015}, Liu \emph{et al.} investigated the minimization of on-grid energy consumption from both the space and time dimensions, and developed a low-complexity offline algorithm based on non-causal information as well as several heuristic online algorithms based on only causal information. However, both \cite{Jie2018} and \cite{Dantong2015} rely on the assumption that the uncertainties follow some well-known probability distributions such as Poisson distribution. They are not suitable for the scenario where the practical probability distributions disagree with the pre-assumed statistical models. 

{To facilitate joint optimization of energy and communication resource allocation under distribution free models, Lyapunov optimization has been widely used to provide bounded performance guarantees of resource allocation under all possible realizations of uncertainties \cite{Moritz2011}. It has been applied in wireless networks \cite{Weihua2016}, hybrid energy powered cellular networks \cite{Yuyi2015}, and relay cooperative networks \cite{Yang2018Lyapunov}, etc. Nevertheless, the above-mentioned works mainly focus on one-timescale stochastic models, and cannot be directly applied to solve the two-timescale resource allocation problem addressed in this paper. Moreover, they cannot well handle the large-scale resource allocation problem with massive IIoT devices. Alternating direction method of multipliers (ADMM) enables low-complexity optimization \cite{ZhangDistributed}. However, it cannot be directly applied for the two-timescale resource allocation problem of IIoT due to the coupling between energy resource allocation and communication resource allocation in different timescales and layers.}

\subsection{Contribution}

Motivated by these gaps, we propose a two-timescale resource allocation algorithm for 5G-empowered automated networks in IIoT with hybrid energy supply. 
The main objective is to maximize the long-term network utility via the joint optimization of communication and energy resource allocation under dynamic EH, electricity price, channel state, data arrival, as well as the long-term constraints of queue stability and queuing delay. 
First, we establish both data and energy queues in different timescales. The joint optimization problem is formulated as a long-term reward-plus-penalty problem, in which the network quality of experience (QoE) is taken as the reward while the energy purchasing cost is taken as the penalty. Then, the long-term problem is further converted to a short-term deterministic problem and decomposed into several subproblems in different timescales by leveraging Lyapunov optimization. Next, by opportunistically minimizing the upper bound of drift-minus-utility, the separated energy management, rate control, channel selection and power allocation subproblems are solved sequentially by using the proposed heurist energy scheduling algorithm, ADMM-based low complexity rate control algorithm, and matching-based joint channel selection and power allocation algorithm, respectively.

The main contributions are summarized as follows.

\begin{itemize}
\item \emph{ Large-timescale energy management optimization under dynamic EH and electricity price:} The proposed algorithm decouples the large-timescale energy management optimization from the small-timescale communication resource allocation. The proposed heuristic energy scheduling algorithm dynamically optimizes the utilization of harvested energy and grid energy without requiring any prior knowledge of future EH and electricity prices.
\item \emph{Small-timescale joint optimization of rate control, channel selection, and power allocation:} The proposed ADMM-based low-complexity rate control algorithm decomposes the large-scale optimization problem into a series of subproblems with lower complexity and accelerates the convergence speed via effective coordination of subproblem solutions. The joint optimization of channel selection and power control is transformed into a one-to-many matching problem and solved by a proposed price-based matching algorithm with quota restriction. 
\item \emph{Comprehensive theoretical analysis and performance validation:} We provide a comprehensive theoretical analysis for the proposed algorithm in terms of optimality, convergence, and complexity. Intensive simulation results are conducted under different scenarios to demonstrate its performance gains.
\end{itemize}
\subsection{Organization}
The rest of this paper is organized as follows. System model is described in Section \ref{sec:3}. Problem formulation and transformation are provided in Section \ref{sec:Formulated Problem}. Section \ref{sec:4} elaborates the proposed two-timescale resource allocation algorithm. A comprehensive property analysis is provided in Section \ref{sec:6}. Numerical results and analysis are introduced in
Section \ref{sec:7}. Finally, the conclusion is summarized in Section  \ref{sec:8}.

%%%%%%%%%%%%%%%%%%%%%%%%%%%%%%%%%%%%%%%%%%%%%%%%%%%%%%%%%%%%%%%%%%%%%%%%%%%%%%%%%%%%%%%%%%%%%%%%%%%%%%%%%%%%%%%%%%%%%%%%%%%%%%%%%%%%%%
%%%%%%%%%%%%%%%%%%%%%%%%%%%%%%%%%%%%%%%%%%%%%%%%%%%%%%%%%%%%%%%%%%%%%%%%%%%%%%%%%%%%%%%%%%%%%%%%%%%%%%%%%%%%%%%%%%%%%%%%%%%%%%%%%%%%%%
%%%%%%%%%%%%%%%%%%%%%%%%%%%%%%%%%%%%%%%%%%%%%%%%%%%%%%%%%%%%%%%%%%%%%%%%%%%%%%%%%%%%%%%%%%%%%%%%%%%%%%%%%%%%%%%%%%%%%%%%%%%%%%%%%%%%%%
\section{System model}
\label{sec:3}
%\textbf{merge Fig. 1 and Fig. 2 into one figure as shown in the paper of Ying Cui}.

%We consider the downlink scenario of a  hybrid energy powered cellular network for IIoT as shown in Fig. \ref{fig:1}. The BS provides wireless connection and data transmission for the IIoT devices within its coverage. It is connected with a rechargeable battery, which supplements energy by either harvesting energy from external renewable energy sources, or purchasing energy from the power grid. The energy supply volatility caused by intermittent renewable energy sources is compensated by the reliable grid power. In addition, the BS maintains a data buffer to store the burst traffic flow towards the IIoT devices. In the following, the models of timescale difference, data queue, and energy queue are introduced in details.

\begin{figure}[h]
\begin{center}
\includegraphics[width=0.45\textwidth]{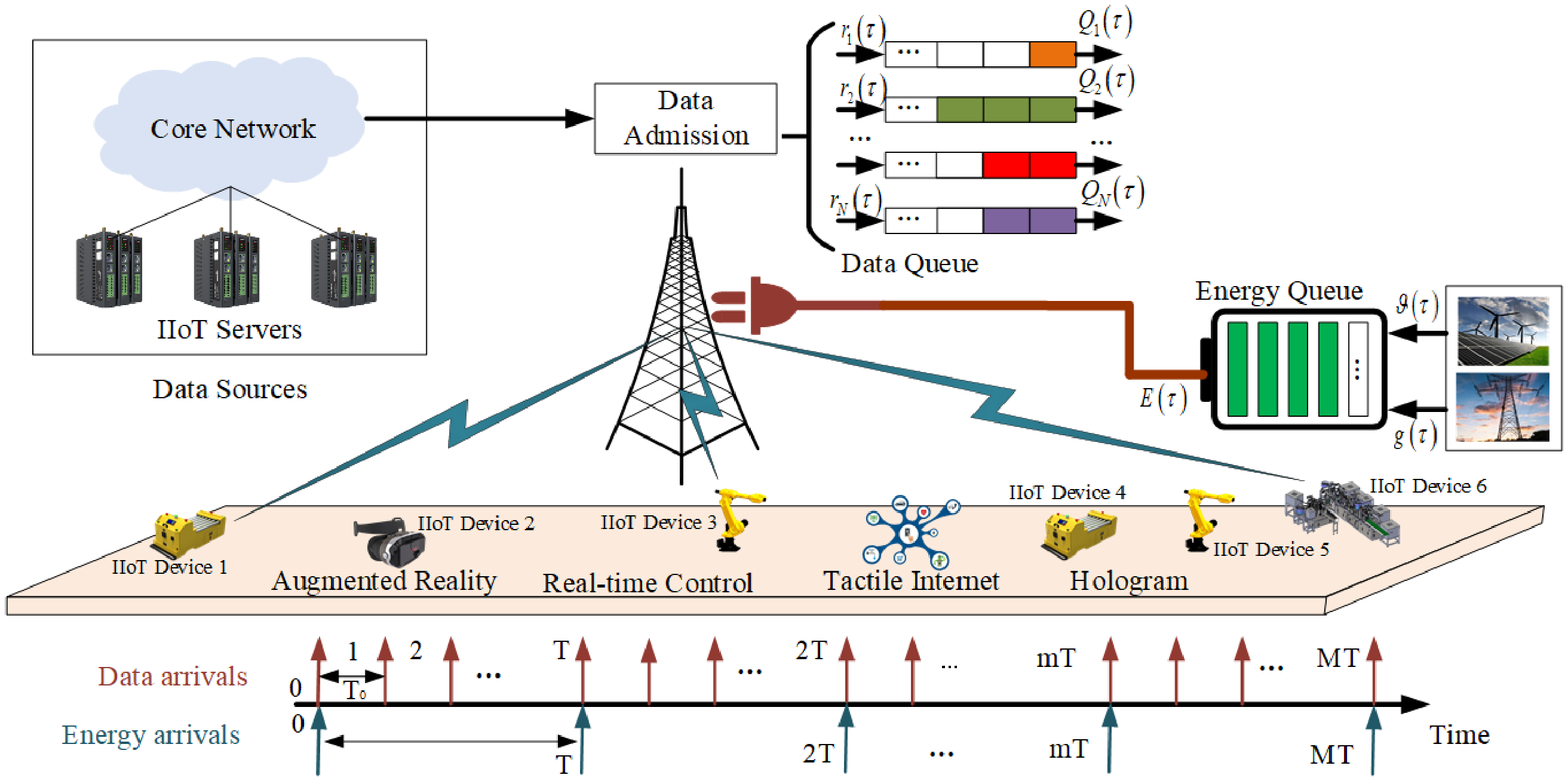}
{\caption{Automated networks with hybrid energy supply for IIoT applications.} \label{system_model}}
\end{center}
\end{figure}

The specific scenario is shown in Fig. \ref{system_model}. The BS provides wireless connection and data transmission for IIoT devices within its coverage. It is connected with a rechargeable battery, which supplements energy by either harvesting energy from external renewable energy sources, or purchasing energy from the power grid. The energy supply volatility caused by intermittent renewable energy sources is compensated by the reliable grid power. We mainly focus on downlink transmission from BS to devices. The reason is that some emerging IIoT applications such as tactile Internet \cite{add4}, augmented reality \cite{add5}, real-time control \cite{add6}, and hologram \cite{add7} impose stringent requirements on downlink data transmission. In such a downlink scenario, the data traffic source is IIoT application servers. The data admitted by the BS are firstly stored in a buffer before transmission, and then are delivered from the BS to the IIoT devices. The IIoT downlink data transmission model has also been adopted and studied in \cite{add1} and\cite{add9}. In the following, the system models are introduced. The key notations are summarized in Table \ref{tabno}.

\begin{table*}[h]
\renewcommand\arraystretch{0.8}
\caption{Summary of Notations.}
\label{tabno}
\begin{center}
\begin{tabular}{|p{35pt}|p{160pt}|p{35pt}|p{160pt}|}
\hline
\textbf{Notations}&\textbf{Definition}&\textbf{Notations}&\textbf{Definition} \\
\hline
$T_0$ & data slot duration & $\tau$ & data slot index \\
\hline
$M$ & number of energy frames & $T$ & number of data slots \\
\hline
$\mathcal{N}$ & set of IIoT devices & $\mathcal{K}$ & set of channels \\
\hline
$r_n(\tau)$ &{amount of data admitted at the BS's network layer for device $n$ at data slot $\tau$} & $p_{n,k}(\tau)$ & {transmission power allocated to device $n$ over channel $k$ at
data slot $\tau$}\\
\hline
$\chi_n$ & {priority of $r_n(\tau)$ to the QoE of device $n$} & $\sigma^2$ & Gaussian white noise power \\
\hline
$x_{n,k}(\tau)$ & {channel selection index for device $n$ over channel $k$ at data slot $\tau$} & $\gamma_{n,k}(\tau)$ & {downlink SNR of device $n$ over channel $k$ at data slot $\tau$} \\
\hline
$U_n(\tau)$ &{QoE for device $n$ at data slot $\tau$} & $h_{n,k}(\tau)$ & {channel gain between the BS and device $n$ over channel $k$ at data slot $\tau$} \\
\hline
$v_{n}(\tau)$ & {downlink transmission rate from the BS to device $n$ at data slot $\tau$} & $W_{k}(\tau)$ & {bandwidth of channel $k$ in data slot $\tau$} \\
\hline
$Q_{n}(\tau)$ & {data backlog of queue $n$ at data slot $\tau$} &$E_{n}(\tau)$ & {energy queue backlog at data slot $\tau$} \\
\hline
$d_{n}$ & {time-average downlink queuing delay of the $n$-th queue} & $d_{n}^*$ & {upper bound of downlink queuing for device $n$} \\
\hline
$\vartheta(\tau)$ & {harvested energy at data slot $\tau$} & $\phi(\tau) $ & {upper bound of $\vartheta(\tau)$ at data slot $\tau$} \\
\hline 
$p_c(\tau)$ & {total amount of energy consumed by the BS at data slot $\tau$} & $g(\tau)$ & {amount of energy purchased from the power grid at data slot $\tau$}\\
\hline
$g_{\max}$ & {upper bound of $g(\tau)$ }& $\beta$ & {whight between energy cost and QoE} \\
\hline 
${\bf{r}}\left( \tau\right)$ & {set of rate control optimization variables at data slot $\tau$} & ${\bf{x}}\left( \tau\right)$ & {set of channel selection optimization variables at data slot $\tau$} \\
\hline 
${\bf{p}}\left( \tau \right)$ & {set of power allocation optimization variables at data slot $\tau$} & $\lambda_1, \lambda_2$ & {Lagrange multiplier corresponding to $C_5$ and $C_6$} \\
\hline 
$\eta(\tau)$ & {electricity price of grid power at data slot $\tau$} & $\kappa(\tau)$ & { electricity price of harvested energy at data slot $\tau$} \\
\hline 
$p_{n,k}^{\max}$ & {maximum transmission power of device $n$ over channel $k$} & $R_{\max}$ & {maximum instantaneous rate of data admission} \\
\hline 
$E_{\max}$ & {battery capacity} &$y$ & Lagrange multiplier in ADMM \\
\hline
$\rho$ & penalty parameter in ADMM &$\mu$ & scaled dual variable in ADMM \\
\hline
$R$ & residual parameter in ADMM &$\epsilon^{pri}, \epsilon^{dual}$ & feasibility tolerances in ADMM\\
\hline
$\mathcal{Q}(\tau)$ & {set of $N$ data queues at data slot $\tau$} & $\mathcal{H}(\tau)$ & {set of $\mathcal{Q}(\tau)$ and $E(\tau)$ at data slot $\tau$} \\
\hline 
$q$ & {maximum number of channels that can be allocated to each device} & $\mathcal{F}$ & set of devices’ preference lists \\
\hline
$\varphi$ & one-to-many matching & $\Lambda_k$ & virtual price of channel $k$ \\
\hline 
\end{tabular}
\end{center}
\end{table*}
\subsection{The Model of Timescale Difference}
\label{sec:31}
Fig. \ref{system_model} shows the timescale difference between data arrival and energy arrival. The two-timescale model proposed in \cite{Cui2015Grid, Zhang2019Two} is adopted, where the continuous time dimension is partitioned into successive identical data slots with duration $T_0$, which is indexed by $\tau=1, 2, \cdots, MT$. Since energy arrival changes much slower than data arrival, we can assume that energy arrival remains constant during $T$ ($T>>1$) data slots \cite{Cui2015Grid}. Therefore, $T$ data slots are grouped as an energy frame with duration of $T T_0$ seconds, which is indexed by $m=1, 2, \cdots, M$.
\subsection{The Model of Data Queue}
\label{sec:32}
Let ${\cal N} = \left\{ {1,2, \cdots ,n, \cdots ,N} \right\}$ and ${\cal K} = \left\{ {1,2, \cdots ,k, \cdots, K} \right\}$ denote the sets of IIoT devices and channels, respectively. Let $r_{n}\left(\tau\right)$ denote the amount of data arriving at the BS's network layer per second for device $n$ at data slot $\tau$, which is firstly stored in the buffer on BS before transmission. The experience of device towards service quality is characterized by QoE \cite{Tran2014QoE}, where the QoE of device $n$ is positively related to the amount of admitted data, which is given by
\begin{align}\label{eq1}
{U_n}\left( \tau  \right) = {\chi_n}\log_{2} \left[ {1 + {r_n}\left( \tau  \right)} \right].
\end{align}

Here, $\chi_n$ is indicates the importance or priority of ${r_n}\left( \tau  \right)$ to the QoE of
device $n$. The logarithmic function is utilized to represent the downtrend of the marginal increment of QoE. Some other works have also adopted logarithmic function-based utility \cite{Guo2017Quality, Jia2017Barrier}. {The QoE model presented in this work is also adopted in \cite{add1, add14}. Compared with directly optimizing physical-layer QoS metrics such as data rate, throughput, and delay, the proposed QoE model can achieve cross-layer optimization between the network layer rate control and the physical layer throughput performance. Besides, the proposed QoE model can also meet the differentiated service requirements through different priority settings. We also consider other QoS performance metrics of queue stability, rate control, downlink queuing delay, and instantaneous delay in the optimization constraints, which means that the communication is not ``best effort”.} 

Let ${x_{n,k}}\left( \tau  \right) \in \left\{ {0,1} \right\}$ be the channel selection index. When ${x_{n,k}\left( \tau  \right)}=1$, the downlink signal to noise ratio (SNR) of device $n$ over channel $k$ is given by
\begin{align}\label{eq2}
{\gamma_{n,k}}\left( \tau  \right) = \frac{{{p_{n,k}}\left( \tau  \right){h_{n,k}}\left( \tau  \right)}}{{{\sigma ^2}}},
\end{align}
where ${p_{n,k}}\left( \tau  \right)$ is the transmission power allocated to device $n$ over channel $k$. ${h_{n,k}}\left( \tau  \right)$ is the channel gain. ${\sigma^2}$ is the Gaussian white noise power.

Then, the downlink transmission rate ${v_n}\left( \tau  \right)$ from the BS to device  $n$ can be derived according to the Shannon-Hartley theorem \cite{Shannon1948}, i.e.,
\begin{align}\label{eq3}
{v_n}\left( {\tau } \right) =\displaystyle\sum\limits_{k = 1}^K {x_{n,k}}\left( \tau  \right){W_k}\left( \tau  \right)\log_{2} \left[ {1 + {\gamma_{n,k}}\left( \tau  \right)} \right],
\end{align}
where ${W_k}\left( \tau  \right)$ denotes the bandwidth of channel $k$.

The data buffered at the BS towards each IIoT device can be regarded as a data queue. Denote the data queue related to device $n$ as queue $n$, where ${r_n}\left( {\tau} \right){T_0}$ and ${v_n}\left( {\tau} \right){T_0}$ can be regarded as the data input and data output, respectively. Particularly,
${r_n}\left( {\tau} \right){T_0}$ indicates how much data related to device $n$ should be sent to the BS in the view of the network layer at data slot $\tau$ and ${v_n}\left( {\tau} \right){T_0}$ indicates how much data should be sent from the BS to device $n$ via wireless link in the physical layer at data slot $\tau$. Let ${Q_n}\left( {\tau } \right)$ denote the data backlog of queue $n$ at data slot $\tau$, which is evolved as
\begin{align}
\label{eqq}
{Q_n}\left( {\tau  + 1} \right) = \max \left[ {{Q_n}\left( \tau  \right) - {v_n}\left( \tau  \right){T_0},0} \right] + {r_n}\left( \tau  \right){T_0}.
\end{align}

${Q_n}\left( \tau  \right)$ is  mean rate stable \cite{Neely2010Stochastic} if $\mathop {\lim }\limits_{{\tau} \to \infty } \displaystyle\frac{{\mathbb{E}\left[ {\left| {{Q_n}\left( \tau  \right)} \right|} \right]}}{{\tau}}{\rm{ = }}0$, which implies that the time-average data output is greater than or equal to the time-average data input, i.e.,
\begin{align}
\mathop {\lim }\limits_{M \to \infty } \frac{1}{{MT}}\sum\limits_{\tau  = 1}^{MT} {{v_n}\left( \tau  \right)}  \ge \mathop {\lim }\limits_{M\to \infty } \frac{1}{{MT}}\sum\limits_{\tau  = 1}^{MT} {{r_n}\left( \tau  \right)}.
\end{align}

The network is considered to be mean rate stable if $\mathop {\lim }\limits_{{\tau} \to \infty } \displaystyle\frac{{\mathbb{E}\left[ {\left| {{Q_n}\left( \tau  \right)} \right|} \right]}}{{\tau}}{\rm{ = }}0$ holds for any device $n \in \mathcal{N}$.

In addition, the time-average communication delay $d_n$ after the $n$-th queue stabilizations \cite{Massey1997} is given by 
%According to Little's theorem \cite{Little2008Little}, the time-average communication delay $d_n$ is given by%
\begin{align}
\label{eqdelay}
&{d_n}= \nonumber \\ 
& \dfrac{{\mathop {\lim }\limits_{M \to \infty } \dfrac{1}{{MT}}\displaystyle\sum\limits_{\tau  = 1}^{MT} {{r_n}\left( \tau  \right)} }}{{\left\{\mathop {\lim }\limits_{M \to \infty } \dfrac{1}{{MT}}\displaystyle\sum\limits_{\tau  = 1}^{MT} {{v_n}\left( \tau  \right)}\right\}
\left\{ {\mathop {\lim }\limits_{ M\to \infty } \dfrac{1}{{MT}}\displaystyle\sum\limits_{\tau  = 1}^{MT} {\left[ {{v_n}\left( \tau  \right) - {r_n}\left( \tau  \right)} \right]} } \right\}}} \nonumber \\ &\le d _n^ *,
\end{align}
where $d_n^ *$ is the upper bound of delay for device $n$.

\emph{{Remark 1:}} Since data arrival and CSI vary across different data slots, the BS has to schedule the values of ${r_n}\left( {\tau} \right)$, ${x_{n,k}}\left( \tau \right)$, and ${p_{n,k}}\left( \tau \right)$ for each device $n \in \mathcal{N}$.

%%%%%%%%%%%%%%%%%%%%%%%%%%%%%%%%%%%%%%%%%%%%%%%%%%%%%%%%%%%%%%%%%%%%%%%%%%%%%%%%%%%%%%%%%%%%%%%%%%%%%%%%%%%%%%%
\subsection{The Model of Energy Queue}
\label{sec:33}

The BS can either exploit renewable energy or purchase grid power. Denote the harvested energy at data slot $\tau$ as $\vartheta \left( \tau  \right)$, which satisfies the following EH constraint
\begin{align}
0 \le \vartheta \left( \tau  \right) \le \Phi \left( \tau  \right),
\end{align}
where $\Phi \left( \tau  \right)$ denotes the upper bound of harvested energy. Denote ${g}\left( \tau  \right)$ as the amount of energy purchased from the power grid, which is bounded by ${g_{\max }}$.

\emph{Remark 2:} Notably, the EH process and electricity price vary  much slower than data arrival and channel fading. The latter two change at every data slot, while the former two change at every energy frame, i.e., every $T$ data slots. In order to accomplish stable power supply, the grid energy is expected as a supplement of the renewable energy. As a result, the BS has to schedule $\vartheta \left( \tau  \right)$ and ${g}\left( \tau  \right)$ on the same time scale.

 Denote ${p_{c}}\left(\tau \right)$ as the total amount of energy consumed by the BS at data slot $\tau$, which is given by
\begin{align}
{p_{c}}\left(\tau \right)=\displaystyle\sum\limits_{n=1}^N{\displaystyle\sum\limits_{k = 1}^K {{x_{n,k}}\left( \tau  \right){p_{n,k}}\left( \tau\right)}}{T_0}.
\end{align}

The battery state of the BS is regarded as an energy queue and the energy queue backlog $E\left(\tau\right)$ is evolved as
\begin{align}\label{eqe}
E\left({\tau+1}\right)=\max\left[{E\left(\tau\right)-{p_{c}}\left(\tau\right),0}\right]+{g}\left(\tau\right)+\vartheta\left(\tau \right).
\end{align}

{Similarly, $E(\tau)$ is mean rate stable as long as $\mathop {\lim }\limits_{{\tau} \to \infty } \displaystyle\frac{{\mathbb{E}\left[ {\left| {{E}\left( \tau \right)} \right|} \right]}}{{\tau}}{\rm{ = }}0$ holds. } Since we focus on the downlink scenario, there are $N$ data queues corresponding to the data of $N$ devices stored in the buffer of the BS before downlink transmission, and one energy queue corresponding to the energy state of the BS. In comparison, the models in \cite{zhangyanTWC, LUOenergy} mainly focus on the uplink scenario where each device holds a data queue and an energy queue. According to the causality constraint, i.e., ${p_{c}}\left( \tau \right) \le E\left( \tau \right)$,
the consumed energy cannot exceed the currently available energy in the battery. On the other hand, the energy queue backlog is also limited by the battery capacity ${E_{\max }}$, i.e.,
\begin{align}\label{eq10}
E\left( \tau \right) + {g}\left( \tau \right) + \vartheta \left( \tau \right) \le {E_{\max }}.
\end{align}

\section{Problem Formulation and Problem Transformation}
\label{sec:Formulated Problem}
{
In this section, we first introduce the problem formulation. Then, the Lyapunov optimization-based problem transformation is elaborated. }

\subsection{ Problem Formulation }{
In this paper, we aim at maximizing the long-term QoE performance of the overall network while minimizing the energy cost.} The objective function is defined as a weighted sum of QoE and energy cost, which is given by
\begin{align}
f\left( \tau \right) = \displaystyle\sum\limits_{n = 1}^N {{U_n}\left( \tau \right)} - \beta \left[ \eta \left( \tau \right){g}\left( \tau \right)+\kappa\left(\tau \right) \vartheta \left( \tau \right) \right],
\end{align}
where $\eta \left( \tau \right)$ and $\kappa\left(\tau \right)$ are the electricity prices of grid power and harvested energy. $\beta $ is a parameter used to balance the tradeoff between energy cost and QoE. We adopt a real-time electricity price model which varies in the same timescale of energy harvesting. Similar electricity price timescale has also been adopted in other works \cite{Talhar_2020, Muratori_2016}.

Denote ${\bf{r}}\left( \tau\right) = \left\{ {{r_n}\left( \tau  \right)} \right\}$, ${\bf{x}}\left( \tau\right) = \left\{ {{x_{n,k}}\left( \tau  \right)} \right\}$, and ${\bf{p}}\left( \tau \right) = \left\{ {{p_{n,k}}\left( \tau  \right)} \right\}$. The two-timescale resource allocation problem with long-term optimization objective is formulated as
\begin{align}\label{eq12}
\begin{array}{l}
{{\rm{\bf{P1}:}}}\mathop {{\rm{maximize}}}\limits_{
{g\left( \tau \right),\vartheta \left( \tau \right),{\bf{r}}\left( \tau \right),{\bf{x}}\left( \tau \right),{{\bf{p}}}\left( \tau \right)}
} \mathop {\lim }\limits_{{M} \to \infty } \dfrac{1}{{MT} }\displaystyle\sum\limits_{\tau = 1}^{{MT} } {f\left( \tau \right)} \\
{\rm{s.t.}}\\
{\rm{C_1:}}{\kern 1pt}{\kern 1pt}0 \le \vartheta \left( \tau \right) \le \Phi \left( \tau \right) ,\forall \tau,\\
{\rm{C_2:}}{\kern 1pt}{\kern 1pt}0 \le g\left( \tau \right) \le {g_{\max }} ,\forall \tau,\\
{\rm{C_3:}}{\kern 1pt}{\kern 1pt}0\le E\left( \tau \right) + {g}\left( \tau \right) + \vartheta \left( \tau \right) \le {E_{\max }} ,\forall \tau, \\
{\rm{C_4:}}{\kern 1pt}{\kern 1pt}0 \le {p_{c}}\left( \tau \right) \le E\left( \tau \right) ,\forall \tau, \\
{\rm{C_5:}}{\kern 1pt}{\kern 1pt}0 \le {p_{n,k}}\left( \tau \right) \le p_{n,k}^{\max }, \forall k, \forall n, \forall \tau, \\
{\rm{C_6:}}{\kern 1pt}{\kern 1pt}{x_{n,k}}\left( \tau \right) \in \left\{ {0,1} \right\}, \forall k,\forall n,\forall \tau,\\
{\rm{C_7:}}{\kern 1pt}{\kern 1pt}\displaystyle\sum\limits_{k = 1}^K {{x_{n,k}}\left( \tau \right)} \le q, \forall n,\forall \tau,\\
{\rm{C_8:}}{\kern 1pt}{\kern 1pt}\displaystyle\sum\limits_{n = 1}^N {{x_{n,k}}\left( \tau \right)} \le 1, \forall k,\forall \tau,\\
{\rm{C_9:}}{\kern 1pt}{\kern 1pt}0 \le \displaystyle\sum\limits_{n = 1}^N {{r_n}\left( \tau \right)} \le {R_{\max}},\forall \tau, \\
{\rm{C_{10}:}}{\kern 1pt}{\kern 1pt}d_n\le d_n^*, \forall n,\\
{\rm{C_{11}:}}{\kern 1pt}{\kern 1pt} E,{\kern 1pt} {\kern 1pt} Q_n, \forall n, {\kern 1pt} {\kern 1pt}\rm{are} {\kern 1pt} {\kern 1pt} \rm{mean} {\kern 1pt}{\kern 1pt}\rm{rate}{\kern 1pt}{\kern 1pt} \rm{stable}.
\end{array}
\end{align}

Here, ${\rm{C_1}}$ and ${\rm{C_2}}$ denote the upper bounds of harvested energy and purchased energy, respectively. ${\rm{C_3}}$ is the battery capacity constraint. ${\rm{C_4}}$ is the energy causality constraint.  ${\rm{C_5}}$ is the instantaneous constraint of transmission power, and $p_{n,k}^{\max}$ is the maximum transmission power.  ${\rm{C_6}}-{\rm{C_8}}$ denote the channel selection constraints, i.e., each channel could be only used by one device and one device could use at most $q$ channels. $q$ also indicates the quota of channel for device, i.e., the maximum number of channels that can be allocated to each device. ${\rm{C_9}}$ is the instantaneous rate control constraint of the overall network, and $R_{\max}$ is the maximum instantaneous rate of data arrival. $\rm{C_{10}}$ is the time-average delay constraint. $\rm{C_{11}}$ denotes the stability constraints of data queue and energy queue.

There exist some difficulties when solving $\mathbf{P1}$. First, the prior knowledge of future CSI, data arrival, energy arrival and electricity price is unknown. Second, it involves resource allocation in different timescales, i.e., rate control, channel selection, and power allocation have to be jointly optimized every data slot, while energy management has to be optimized every energy frame. Third, $\mathbf{P1}$ is a mixed integer nonlinear problem (MINP), which is NP-hard due to the coupling between long-term constraints and short-term optimization objectives \cite{NPHard2019, NPHardNew}. Therefore, it is more complicated than traditional mixed integer nonlinear optimization problems. Last but not least, the sum-rate constraint ${\rm{C_9}}$ raises complexity issues as the problem dimension increases significantly with the number of IIoT devices.
\subsection{Lyapunov Optimization-based Problem Transformation }{
Lyapunov optimization is introduced to transform the long-term optimization problem into a series of single-frame optimization subproblems, which are further decomposed over two timescales.} Denote ${\cal Q}\left( \tau \right)=\left[ {Q_1}\left( \tau \right),{Q_2}\left( \tau \right), \cdots ,{Q_N}\left( \tau \right)\right]$. Let ${\cal H}\left( \tau \right) = \left[ {{\cal Q}\left( \tau \right),E\left( \tau \right)} \right]$ be a concatenated vector of queue states. Subsequently, based on\cite{Zhang2019Two}, the Lyapunov function is defined as 
\begin{align}\label{eq13}
L\left( \tau  \right) = \dfrac{1}{2}\left\{ {\displaystyle\sum\limits_{n = 1}^N {Q_n^2\left( \tau  \right)}  + {{\left[ {{E_{\max }} - E\left( \tau  \right)} \right]}^2}} \right\}.
\end{align}

The Lyapunov drift over $T$ data slots conditioned on the states of both data and energy queues is given by
\begin{align}\label{eq14}
{\Delta_T}\left( \tau  \right) = \mathbb{E}\left[ {L\left( {\tau  + T} \right) - L\left( \tau  \right)\left| {{\cal H}\left( \tau  \right)} \right.} \right].
\end{align}

Accordingly, the drift-minus-utility (DMU) function is defined as
\begin{align}\label{eq15}
D\left[ {{\cal H}\left( \tau  \right)} \right]{\rm{ = }}\mathbb{E}\left[ {{\Delta _T}\left( \tau  \right) - Vf\left( \tau  \right)\left| {{\cal H}\left( \tau  \right)} \right.} \right],
\end{align}
where $V$ is a tunable weight which represents the relative importance of ``utility maximization'' compared with ``queue stability''.

Considering the timescale difference among energy management, rate control, channel selection, and power allocation, the upper bound of $D\left[ {{\cal H}\left( \tau  \right)} \right]$ is derived based on the following theorem.
%%%%%%%%%%%%%%%
%%%%%%%%%%%

\begin{theorem}\label{Theorem1}
The DMU function $D\left[ {{\cal H}\left( \tau  \right)} \right]$ is upper bounded by
\begin{align}
D\left[ {{\cal H}\left( \tau  \right)} \right]
\le& \displaystyle\frac{1}{2}\left[ {B + \displaystyle\frac{{\left( {T - 1} \right)}}{2}{{\left( {{{g}_{\max }} + {\vartheta _{\max }}} \right)}^2}} \right]T\notag\\
+& {\mathbb{E}}\bigg\{ {D_1}\left[ {(m-1)T+1} \right]\notag\\
{\rm{ + }}&\displaystyle\sum\limits_{\tau  = (m-1)T+1}^{mT} {\left[{D_2}\left( \tau  \right)-{{D_3}\left( \tau  \right)}\right]}\bigg\} \label{eq18},
\end{align}
where
\begin{align}
\label{eq17}
&B= N\left( {{r_{\max }^2} + {v_{\max }^2}} \right){T_0^2} + {E_{\max }^2} + {\left( {{{g}_{\max }} + {\vartheta _{\max }}} \right)^2}, \nonumber\\
&{\cal E}\left[ {(m-1)T+1} \right]= {E_{\max }} - E\left[ {(m-1)T+1} \right],\nonumber\\
&{D_1}\left[ {(m-1)T+1} \right]= VT\beta \eta \left[ {(m-1)T+1}\right] {g}\left[ {(m-1)T+1} \right] \nonumber \\ 
&- {\cal E}\left[ {(m-1)T+1} \right]{g}\left[ {(m-1)T+1} \right] \nonumber \\
&+ VT\beta \kappa \left[ {(m-1)T+1} \right]\vartheta \left[ {(m-1)T+1} \right] \nonumber \\
&-{\cal E}\left[ {(m-1)T+1} \right] \vartheta \left[ {(m-1)T+1} \right]
, \nonumber\\
&{D_2}\left( \tau  \right)=\displaystyle\sum\limits_{n = 1}^N {\left[ { {{Q_n}\left( \tau  \right){r_n}\left( \tau  \right){T_0}}  - V{U_n}\left( \tau  \right)} \right]}, \nonumber\\
&{D_3}\left( \tau  \right)= \displaystyle\sum\limits_{n = 1}^N {\displaystyle\sum\limits_{k = 1}^K {{x_{n,k}}} } \left( \tau  \right)\left[ {{Q_n}\left( \tau  \right){v_n}\left( \tau  \right){T_0} - {\cal E}\left( \tau  \right){p_{n,k}}\left( \tau  \right)} \right].
\end{align}
\end{theorem}
\begin{IEEEproof}
{See Appendix \ref{APPA}.}
\end{IEEEproof}

In Theorem \ref{Theorem1}, $B$ is a positive constant. Following Lyapunov optimization, $\mathbf{P1}$ is transformed into opportunistically minimizing the right-hand side of (\ref{eq18}) at each energy frame subject to ${{\rm{C}}_{\rm{1}}}\sim{{\rm{C}}_{\rm{10}}}$. Thus, the long-term stochastic optimization problem $\mathbf{P1}$ is converted into a deterministic  short-term optimization problem, which is given by

\begin{align}
&{{\rm{\bf{P2}:}}}\mathop {{\rm{minimize}}}\limits_{
{g\left[(m-1)T+1\right],\vartheta \left[(m-1)T+1\right],{\bf{r}}\left( \tau  \right),{\bf{x}}\left( \tau  \right),{{\bf{p}}}\left( \tau  \right)}
} {D_1}\left[ {(m-1)T+1} \right] \nonumber \\
&+\displaystyle\sum\limits_{\tau  = (m-1)T+1}^{mT} {\left[{D_2}\left( \tau  \right) - {D_3}\left( \tau  \right)\right]} \nonumber \nonumber \\
&{\rm{s.t.}}{\kern 1pt} {\kern 1pt} {{\rm{C}}_{\rm{1}}}-{{\rm{C}}_{\rm{10}}}. \label{eq12.1}
\end{align}

{It is noted that the first term of $\mathbf{P2}$ involves only the energy management decisions, i.e., $g\left[(m-1)T+1\right]$ and $\vartheta\left[(m-1)T+1\right]$. The second term involves only the rate control decisions, i.e., ${\bf{r}}\left( \tau\right)$. The third term involves only the joint channel selection and power allocation decisions, i.e., ${\bf{x}}\left( \tau\right)$ and ${\bf{p}}\left( \tau \right)$. Therefore, we can further decompose $\mathbf{P2}$ into three subproblems in different timescales, i.e., large-timescale energy management subproblem ${\rm{\bf{P3}}}$, small-timescale rate control subproblem ${\rm{\bf{P4}}}$, and small-timescale joint channel selection and power allocation subproblem ${\rm{\bf{P6}}}$, which are introduced in Section \ref{sec:4}.}
\section{Two-timescale Resource Allocation Optimization}
\label{sec:4}
In this section, we aim to solve above two-timescale optimization subproblems. First, the large-timescale energy management subproblem is solved based on linear programming. Second, ADMM is introduced to solve the large-scale rate control problem with the sum-rate constraint. Then, a joint channel selection and power allocation algorithm is developed by leveraging price-based one-to-many matching. The proposed two-timescale resource allocation algorithm is summarized in Algorithm \ref{Lyapunov optimization}.
\begin{algorithm}[bt]
\caption{Two-timescale Resource Allocation Algorithm}
\label{Lyapunov optimization}
\begin{algorithmic}[1]
\STATE \textbf{Input:} $N$, $K$, $T$, $M$, $\left\{h_{n,k}(\tau)\right\}$, ${E_{\max}}$, ${g_{\max}}$, $\left\{p_{n,k}^{\max }\right\}$, ${R_{\max}}$, $q$.
\STATE \textbf{Output:} ${\bf{g}}^ *$, ${\bm{\vartheta} }^ * $, ${\bf{r}}^ * $, ${\bf{x}}^ *$, ${\bf{p}}^ *$.
\STATE \textbf{Initialize:} $\left\{{Q_n}\left( 1 \right)\right\}$, $E\left( 1 \right)$.
\FOR {$m=1:M$}
\STATE \textbf{Energy management:} Obtain the optimal solution ${g^ * }\left[ {(m-1)T+1} \right]$ and ${\vartheta ^ * }\left[ {(m-1)T+1} \right]$ according to (\ref{eq20}) and (\ref{eq21}).
\FOR {$t=1: T$}
\STATE \textbf{Rate control:} Obtain the optimal solution $r_n^ * \left[{\left( {m -1} \right)T+t}\right], \forall n \in {\cal{N}}$, by Algorithm 2.
\STATE \textbf{Joint channel selection and power allocation:} Obtain the optimal solution $x_{n,k}^ * \left[{\left( {m -1} \right)T+t}\right]$, by the proposed one-to-many matching. Obtain the optimal solution $p_{n,k}^ * \left[{\left( {m -1} \right)T+t}\right]$ according to (\ref{opp}).
\STATE Update all the data queues ${Q_n}\left( \tau \right), \forall n \in {\cal{N}}$, and the energy queue $E\left(\tau\right)$ according to (\ref{eqq}) and (\ref{eqe}).
\ENDFOR
\ENDFOR
\end{algorithmic}
\end{algorithm}
\subsection{ Large-timescale Energy Management Based on Linear Programming }
Accordingly, BS schedules the harvested energy and purchased energy every $T$ data slots. To minimize $D_1\left[ {(m-1)T+1} \right]$, $\forall m \in \{1,2,\dots,M\}$, we solve the following energy management subproblem
\begin{align}\label{eq22}
\begin{array}{l}
{{\rm{\bf{P3}:}}}\mathop {{\rm{minimize}}}\limits_{g\left[ {(m-1)T+1} \right],\vartheta \left[ {(m-1)T+1} \right]} {\rm{ }}{D_1}\left[ {(m-1)T+1} \right]\\
{\rm{s.t.{\kern 1pt}{\kern 1pt}C_1, C_2, C_3}},\tau=(m-1)T+1.
\end{array}
\end{align}

Under the condition that the price of harvested energy is lower than that of grid power, i.e., $\kappa \left[ {(m-1)T+1} \right] <\eta \left[ {(m-1)T+1} \right]$, minimizing ${D_1}\left[ {(m-1)T+1} \right]$ is equivalent to using harvested energy as much as possible. However, the available amount of harvested energy is limited by the upper bounds of both the harvested energy $\Phi \left[ {(m-1)T+1} \right]$ and the remaining battery capacity ${\cal{E}} \left[(m-1)T+1\right]$. Therefore, the optimal scheduling policy of harvested energy is derived as
\begin{align}
\label{eq20}
{\vartheta ^ * } &[{(m-1)T+1}]{\rm{ = }} \nonumber \\
&\min \left\{ {\Phi \left[ {(m-1)T+1} \right],{{\cal{E}} \left[(m-1)T+1\right]}} \right\}. 
\end{align} 

Taking ${\vartheta^*}\left[{(m-1)T+1}\right]$ into ${D_1}\left[{(m-1)T+1}\right]$, the optimal amount of purchased energy is derived as
\begin{align}
\label{eq21}
\begin{array}{l}
{g^*}\left[ {(m-1)T+1} \right]\\ =
\left\{ {\begin{array}{*{20}{c}}
{\min \left\{ {{\cal{E}}\left[ {(m-1)T+1} \right] - {\vartheta ^*}\left[ {(m-1)T+1} \right],{g_{\max }}} \right\}},\\
 {{\rm{if}}\ {\rm{ }} \Psi \left[ {(m - 1)T + 1} \right] < 0},\\
0,{{\rm{otherwise}}},
\end{array}} \right.
\end{array}
\end{align} 
where 
\begin{align}
\Psi &\left[ {(m - 1)T + 1} \right]{\rm{ = }} \nonumber \\
&VT\beta \eta \left[ {(m - 1)T + 1} \right] - {\cal E}\left[ {(m - 1)T + 1} \right]. 
\end{align} 

{From (\ref{eq20}), we could find the optimization of ${\vartheta^*}\left[{(m-1)T+1}\right]$ does not depend on the price of harvested energy. From (\ref{eq21}), we could find that the judgment conditions of ${g^*}\left[ {(m-1)T+1} \right]$, i.e., $\Psi\left[{(m-1)T+1}\right]$, depend on the price of grid power $\eta \left[ {(m - 1)T + 1}\right]$. However, the optimal values ${g^*}\left[{(m-1)T+1}\right]$ under the condition $\Psi \left[ {(m - 1)T + 1} \right]<0$ or the condition $\Psi \left[ {(m - 1)T + 1} \right]\geq 0$ are independent of the price of grid power. } In addition, it also can be found that they are optimized every energy frame, i.e., $T$ data slots, while the energy queue length $E\left(\tau\right)$ changes over each data slot. Therefore, the energy scheduling policy only depends on the current energy queue state.
\subsection{Low-complexity Small-timescale Rate Control Algorithm Based on ADMM}
\label{sec:ADMM}
To minimize the second term $D_2\left( \tau\right)$, the following rate control subproblem is solved at $\tau \in \left[ {(m - 1)T + 1,mT} \right]$, $\forall m \in \{1,2,\dots,M\}$, which is given by
\begin{align}\label{eq25}
\begin{array}{l}
{{\rm{\bf{P4}:}}}\mathop {{\rm{minimize}}}\limits_{{\bf{r}}\left(\tau\right)} {\rm{ }}{D_2}\left( \tau \right)\\
{\rm{s.t.{\kern 1pt}{\kern 1pt}C_9}}.
\end{array}
\end{align}

Due to the sum-rate constraint $\rm{C_9}$, the optimization variables of different devices are coupled, and the computational complexity grows enormously as the number of devices increases. When the number of IIoT devices is large, it will take tremendous amount of time to solve the large-scale rate control problem. Thus, we propose an ADMM-based low-complexity algorithm to solve the large-scale rate control subproblem. The major concept is to alternatively update primal and dual variables in an iterative fashion \cite{LiangVirtual}. It can rapidly find the optimal solution in low complexity based on the decomposition-coordination approach.

In order to obtain the optimal solution, we partition the vector of rate control variables into two parts, i.e.,
${\bf{x}_{r}}=[r_{1}\left(\tau\right), r_{2}\left(\tau\right), \cdots, r_{l_{r}}\left(\tau\right)]^T$ and ${\bf{z}_{r}}=[r_{{l_{r}}+1}\left(\tau\right), r_{{l_{r}}+2}\left(\tau\right), \cdots, r_{N}\left(\tau\right)]^T$. {Based on \cite{WangFully}, $\rm{\bf{P4}}$ is rewritten as
\begin{align}
\begin{array}{l}\label{admm1}
{{\rm{\bf{P5}:}}}\mathop {{\rm{minimize}}}\limits_{{\bf{x}}_r,{\bf{z}}_r}{F_{r}}\left( {{\bf{x}}_r} \right) + {G_{r}}\left( {\bf{z}}_r \right)\\
{\rm{s}}{\rm{.t}}{\rm{.}}{\kern 1pt} {\kern 1pt} {{\bf{A}}_r{\bf{x}}_r}+{{\bf{B}}_{r}{\bf{z}}_r}= R_{\max}.
\end{array}
\end{align}}
where ${{\bf{x}}_r} \in {{\mathbb{R}}^{{l_r}\times 1}}$, ${\bf{z}}_r \in {{\mathbb{R}}^{\left( {N- {l_r}} \right)\times 1}}$, ${{\bf{A}}_r} \in {{\mathbb{R}}^{1 \times {l_r}}}$, and ${{\bf{B}}_r} \in {{\mathbb{R}}^{1 \times \left( {N - {l_r}} \right)}}$. ${{\bf{A}}_r}$ and ${{\bf{B}}_r}$ are unit vectors. $F_r({\bf{x}}_r)$ and $G_r({\bf{z}}_r)$ satisfy
\begin{align}
{F_r}\left( {{{\bf{x}}_r}} \right) = &{{\bf{Q}}_1}{{\bf{x}}_r} - {{\bf{V}}_{1,\chi }}{\log _2}\left( {{{\bf{x}}_r}} \right),\\
{G_r}\left( {{{\bf{z}}_r}} \right) = &{{\bf{Q}}_2}{{\bf{z}}_r} - {{\bf{V}}_{2,\chi }}{\log _2}\left( {{{\bf{z}}_r}} \right),
\end{align}
where ${{\bf{Q}}_1} = \left[ {{Q_1}\left( \tau \right),{Q_2}\left( \tau \right), \cdots ,{Q_{{l_r}}}\left( \tau \right)} \right]{T_0}$, ${{\bf{Q}}_2} = \left[ {{Q_{{l_r}+1}}\left( \tau \right),{Q_{{l_r}+2}}\left( \tau \right), \cdots ,{Q_{{N}}}\left( \tau \right)} \right]{T_0}$,
${{\bf{V}}_{1,\chi }} = \left[ {{\chi _1},{\chi _2}, \cdots ,{\chi _{{l_r}}}} \right] V$, and ${{\bf{V}}_{2,\chi }} = \left[ {{\chi _{{l_r} + 1}},{\chi _{{l_r} + 2}}, \cdots ,{\chi _N}} \right] V$.

In this paper, we adopt the scaled ADMM algorithm \cite{ZhangDistributed}, and form the augmented Lagrangian associated with $\mathbf{P5}$ as
\begin{align}
{{\bf{L}}_\rho }\left( {{{\bf{x}}_r},{{\bf{z}}_r},{{y}}} \right) = F_r\left( {{\bf{x}}_r} \right) + G_r\left( {{\bf{z}}_r} \right) + \frac{\rho}{2}\left\| {{{R}} + {{\mu }}} \right\|_2^2 - \frac{\rho }{2}\left\| {{\mu }} \right\|_2^2,
\end{align}
where ${{R}} = {{\bf{A}_r}}{{\bf{x}}_r} + {{\bf{B}_r}}{{\bf{z}}_r} - R_{\max}$ is the residual. ${\rho}>0$ represents the penalty parameter, which is related to the convergence speed of ADMM. Let ${y}$ be the Lagrange multiplier. ${{\mu}}=\dfrac{{y}}{\rho }$ is the scaled dual variables. Then, we can iteratively update both primal and dual variables as
\begin{align}
&{{\bf{x}}_r^{i + 1}} = \arg \min \Big\{ {F_r\left( {\bf{x}}_r^i \right)} \nonumber \\
&\ \ \ \ \ \ \ \ \ \ \ \ \ \ \ {+\dfrac{\rho }{2}\left\| {{{\bf{A}}}_r{\bf{x}}_r^i + {{\bf{B}}}_r{{\bf{z}}_r^i}-R_{\max} + {{{\mu }}^i}} \right\|_2^2} \Big\} \label{eqn_dbl_x}, \\
&\bf{z}_r^{i + 1} = \arg \min \Big\{ G_r\left( {\bf{z}}_r^i \right) \nonumber \\
&\ \ \ \ \ \ \ \ \ \ \ \ \ \ \ +\dfrac{\rho}{2}\left\| {{{\bf{A}}}_r{{\bf{x}}_r^{i + 1}} + {{\bf{B}}}_r{\bf{z}}_r^i- R_{\max} + {{{\mu }}^i}} \right\|_2^2 \Big\}, \label{eqn_dbl_y}\\
&{{{\mu }}^{i + 1}} = {{{\mu }}^i}{\rm{ + }}{{\bf{A}}}_r{{\bf{x}}_r^{i + 1}} + {{\bf{B}}}_r{{\bf{z}}_r^{i + 1}}-R_{\max},\label{eqn_dbl_z}
\end{align}
where $i$ denotes the index of iteration.

Next, based on the optimality conditions \cite{ChenADMM-Based}, the primal residual ${{\bf{R}}_{\bf{p}}}$ and the dual residual ${{\bf{R}}_{\bf{d}}}$ are expressed as
\begin{align}
\label{RP}
{{\bf{R}}_{\bf{p}}}^{i + 1} =& {{\bf{A}}}_r{{\bf{x}}_r^{i + 1}} + {{\bf{B}}}_r{{\bf{z}}_r^{i + 1}}-{R_{\max}},
\end{align}
\begin{align}
\label{RD}
{{\bf{R}}_{\bf{d}}}^{i + 1} = &\rho {\bf{A}}_r^T{{\bf{B}}}_r\left( {{{\bf{z}}_r^{i + 1}} - {{\bf{z}}_r^i}} \right).
\end{align}

The termination criteria is defined as
\begin{align}
{\left\| {{{\bf{R}}_{\bf{p}}}^{i + 1}} \right\|_2}\le \epsilon^{pri} {\kern 1pt} {\kern 1pt}{\rm{and}} {\kern 1pt} {\kern 1pt} {\left\| {{{\bf{R}}_{\bf{d}}}^{i + 1}} \right\|_2} \le \epsilon^{dual},
\end{align}
where $\epsilon^{pri}>0$ and $\epsilon^{dual}>0$ denote feasibility tolerances with respect to primal conditions and dual conditions. Consequently, the ADMM-based low-complexity rate control algorithm is summarized in Algorithm \ref{ADMMoptimization}.

\begin{algorithm}[t]
\caption{ADMM-based Low-complexity Rate Control Algorithm}
\label{ADMMoptimization}
\begin{algorithmic}[1]
\STATE \textbf{Input:} $i$, ${\bf{x}}_r$, ${\bf{z}}_r$, ${{\mu}}$, ${\rho}$, $\epsilon^{pri}$, and $\epsilon^{dual}$.
\STATE \textbf{Output:} ${\bf{x}}_r$, ${\bf{z}}_r$.
\WHILE {${\left\| {{{\bf{R}}_{\bf{p}}}^{i}} \right\|_2} > \epsilon^{pri}$ or ${\left\| {{{\bf{R}}_{\bf{d}}}^{i}} \right\|_2} > \epsilon^{dual}$}
\STATE {Update ${\bf{x}}_r^{i+1} $ according to (\ref{eqn_dbl_x})};
\STATE {Update ${\bf{z}}_r^{i+1}$ according to (\ref{eqn_dbl_y})};
\STATE {Update ${{\mu}}^{i+1}$ according to (\ref{eqn_dbl_z})};
%\STATE Calculate the termination criteria:
\STATE {{Update ${\left\| {{{\bf{R}}_{\bf{p}}}^{i + 1}} \right\|_2}$ and ${\left\| {{{\bf{R}}_{\bf{d}}}^{i + 1}} \right\|_2}$ according to (\ref{RP}) and (\ref{RD})};}
\STATE {Update $i \rightarrow i+1$}.
\ENDWHILE
\end{algorithmic}
\end{algorithm}

\subsection{Small-timescale Joint Channel Selection and Power Allocation Based on One-to-many Matching}
\label{sec:5}

To maximize the third term $D_3\left( \tau\right)$, the following joint channel selection and power allocation subproblem is solved at $\tau \in \left[ {(m - 1)T + 1,mT} \right]$, $\forall m \in \{1,2,\dots,M\}$, which is given by
\begin{align}\label{eq26}
\begin{array}{l}
{{\rm{\bf{P6}:}}}\mathop {{\rm{maximize}}}\limits_{{{\bf{x}}\left(\tau\right),{\bf{p}\left(\tau\right)}}} {\rm{ }}{D_3}\left( \tau \right)\\
{\rm{s.t.{\kern 1pt}{\kern 1pt}{C_4}-{C_8},C_{10}}}.
\end{array}
\end{align}

Different from traditional mobile devices, IIoT has stringent QoS requirements such as delay. Compared with related works \cite{EMM_sun, EDT_2004}, we not only consider the long-term constraint of queuing delay, but also take strict instantaneous downlink queuing delay constraint into consideration. The tight coupling between $r_n(\tau)$ and $v_n(\tau)$ in (6) makes it difficult to transform $C_{10}$. Thus, we tighten the delay constraints over every data slot as
\begin{align}\label{delay2}
\frac{{{r_n}\left( \tau \right)}}{{{v_n}\left( \tau \right)\left[ {{v_n}\left( \tau \right) - {r_n}\left( \tau \right)} \right]}} \le d_n^ *. 
\end{align}

The instantaneous delay is defined as the left term of (24), i.e., $\frac{{{r_n}\left( \tau \right)}}{{{v_n}\left( \tau \right)\left[ {{v_n}\left( \tau \right) - {r_n}\left( \tau \right)} \right]}}$. It only involves short-term variables of slot $\tau$, i.e., $r_n(\tau)$ and $v_n(\tau)$.

The downlink transmission delay is considered implicitly in the downlink queuing delay. First, to reduce the downlink queuing delay for maintaining queue stability, the BS should select channel with higher quality and allocate more transmission power, both of which will reduce the downlink transmission delay. Second, under the queue stability condition, the time-average data output is greater than or equal to the time-average data input based on (5), i.e., the downlink transmission rate has a lower bound. Since the downlink transmission delay is inversely proportional to the downlink transmission rate, it has an upper bound.

Rearranging (\ref{delay2}), we can get
\begin{align}\label{delay3}
f_d\left[ {{v_n}\left( \tau \right)} \right] = {r_n}\left( \tau \right) - d_n^ * {\left[ {{v_n}\left( \tau \right)} \right]^2} + d_n^ * {v_n}\left( \tau \right){r_n}\left( \tau \right) \le 0,
\end{align}
where $f_d\left[ {{v_n}\left( \tau \right)} \right]$ is a one-variable quadratic inequality with respect to $v_n\left(\tau\right)$. Since $f_d\left( 0 \right) = {r_n}\left( \tau \right) > 0$ and $d_n^ *>0$, there exist a positive solution and a negative solution which make the equality in (\ref{delay3}) hold. The positive solution is given by
%there are two zero points for $f_d\left[ {{v_n}\left( \tau \right)} \right]$. The one that is greater than zero is calculated as
\begin{align}
\begin{array}{l}
v_n^ * \left( \tau \right) = \dfrac{{d _n^ * {r_n}\left( \tau \right) + \sqrt {{{\left[ {d_n^ * {r_n}\left( \tau \right)} \right]}^2} + 4{r_n}\left( \tau \right)d _n^ * } }}{{2d _n^ * }}.
\end{array}
\end{align}

Accordingly, the delay constraint could be converted into a transmission capacity constraint as
\begin{align}\label{delay4}
\begin{array}{l}
{\rm{C_{12}:}}{\kern 1pt}{\kern 1pt}{v_n}\left( \tau \right) \ge v_n^ * \left( \tau \right).\\
\end{array}
\end{align}

Replacing $\rm{C_{10}}$ with $\rm{C_{12}}$, $\mathbf{P6}$ is rewritten as
\begin{align}\label{delay5}
\begin{array}{l}
{{\rm{\bf{P7}:}}}\mathop {{\rm{maximize}}}\limits_{{{\bf{x}}\left(\tau\right)},{\bf{p}\left(\tau\right)}} {\rm{ }}{D_3}\left( \tau \right)\\
{\rm{s.t.{\kern 1pt}{\kern 1pt}C_4-C_8, C_{12}}}.
\end{array}
\end{align}

We rearrange $\mathbf{P7}$ as
\begin{align}\label{jcspa}
\begin{array}{l}
\mathop {{\rm{maximize}}}\limits_{{{\bf{x}}\left(\tau\right)},{\bf{p}\left(\tau\right)}} {\rm{ }}\displaystyle\sum\limits_{n = 1}^N {{f_{{D_3}}}\left[ {{x_{n,k}}\left( \tau \right),{p_{n,k}}\left( \tau \right)} \right]}\\
{\rm{s.t.{\kern 1pt}{\kern 1pt}C_4-C_8, C_{12}}},
\end{array}
\end{align}
where ${f_{{D_3}}}\left[ {{x_{n,k}}\left( \tau \right),{p_{n,k}}\left( \tau \right)} \right] = {Q_n}\left( \tau \right){v_n}\left[ {{x_{n,k}}\left( \tau \right),{p_{n,k}}\left( \tau \right)} \right]{T_0} - {\cal E}\left( \tau \right){x_{n,k}}\left( \tau \right){p_{n,k}}\left( \tau \right)$. {
The problem defined in (\ref{jcspa}) is NP-hard due to the coupling between integer variables and continuous variables. To provide a tractable solution, we transfer it into a one-to-many matching problem. Matching theory has been widely adopted in channel selection optimization. In \cite{match1}, Anandkumar \emph{et al.} considered a cognitive medium access model, and proposed a matching-based joint user allocation algorithm to optimize user access strategy. However, it only considers the one-to-one matching scenario where each channel can only be utilized by at most one user. Different from one-to-one matching, we represent the one-to-many matching problem as a triple $\left( {{\cal N},{\cal K},{\cal F}} \right)$. $\mathcal{N}$ and $\mathcal{K}$ represent the sets of matching participants, i.e., devices and channels, respectively. $\mathcal{F}$ denotes the set of devices’ preference lists.  The definition of one-to-many matching is given as follows.} 
{
\begin{definition}
\textbf{(One-to-many matching)} 
$\varphi$ is a one-to-many correspondence mapping from set $\mathcal{N} \cup \mathcal{K}$ onto itself under preference $\mathcal{F}$, i.e., $\varphi (n) \subseteq {\cal K}$, $\forall n \in {\cal N}$. $k \in \varphi \left( n \right)$ means that channel $k$ is matched with device $n$, i.e., ${x_{n,k}} = 1$. $\varphi \left( n \right) = \{n\}^{1\times q}$ represents that device $n$ is not matched with any channel. The quota $q$ represents that at most $q$ channels can be matched to one device simultaneously, while each channel could be only used by at most one device. 
\end{definition}}
{
Taking device $n$ and channel $k$ as an example to explain matching stability. A matching $\varphi$ is blocked if $n$ and $k$ are not matched but both $n$ and $k$ prefer to be matched with each other under $\varphi$. Thus, $n$ and $k$ form a blocking pair for matching $\varphi$, namely that $\left({n,k}\right)$ blocks the matching. We say that matching $\varphi$ is not stable because $n$ and $k$ would prefer to disrupt the matching in order to be matched with each other.}
\begin{definition}
\label{DF1}
\textbf{(Stable matching)} A matching $\varphi$ is stable if there exists no blocking pair.
\end{definition}
{
One-to-many matching problem has been widely studied. In \cite{match2}, Sanguanpuak \emph{et al.} studied the nonorthogonal spectrum assignment problem, and proposed a many-to-one matching-based spectrum sharing algorithm. However, it does not consider the coupling between power allocation and channel selection.  To decouple the coupling between power allocation and channel selection in the transformed one-to-many matching problem, the preference of device towards channel is established based on the optimal power allocation strategy.} { The proposed joint channel selection and power allocation algorithm based on one-to-many matching is summarized in Algorithm \ref{matchinga}, which contains five phases, i.e., initialization, power allocation, preference list construction, proposal and price rising, and matching termination.  When multiple devices compete for the same channel, classical matching approaches solve the matching conflicts by randomly assigning the channels to a device \cite{RM}. In comparison, we propose the price-based matching with quota restriction, where the price of the specific channel, i.e., the matching cost, is increased to force some device to give up this channel. Algorithm \ref{matchinga} is implemented as follows.}
 {\subsubsection{Initialization} Set $\varphi \left( n \right) = \emptyset$, $\Omega=\emptyset$, and ${\Lambda _k} = 0$, $\forall k \in {\cal K}$. $\Omega$ represents the set of channels which receive more than one matching proposal from devices. ${\Lambda _k}$ is the virtual price of channel $k$ used to solve the conflict of matching.
\subsubsection{Power allocation} By temporarily matching each device $n \in {\cal N}$ with each channel $k\in {\cal K}$, when $\varphi \left( n \right) = k$, the maximum value of ${f_{{D_3}}}\left[ {{p_{n,k}}\left( \tau \right)\left| {{x_{n,k}}\left( \tau \right) = 1} \right.} \right]$ can be obtained by solving the following power allocation problem
\begin{align}\label{eq27}
\begin{array}{l}
{{\rm{\bf{P8}:}}}\mathop {{\rm{maximize}}}\limits_{{p_{n,k}}\left( \tau \right)} {\rm{ }}{f_{D_3}}\left[ {p_{n,k}\left( \tau \right)\left| {{x_{n,k}}\left( \tau \right) = 1} \right.} \right]\\
{\rm{s.t.}}{\kern 1pt}{\kern 1pt}{\rm{C_4}, \rm{C_5}}.
\end{array}
\end{align}}
{$\mathbf{P8}$ is a convex optimization problem and can be solved by applying Karush-Kuhn-Tucker (KKT) conditions. The Lagrangian associated with $\mathbf{P8}$ is given by
\begin{align}
{\cal L}\left[ {{p_{n,k}}\left( \tau \right),\lambda } \right] = &- {f_{D_3}}\left[ {p_{n,k}\left( \tau \right)\left| {{x_{n,k}}\left( \tau \right) = 1} \right.} \right] \nonumber \\
&+ \lambda \left[ {{p_{n,k}}\left( \tau \right) - p_{n,k}^{\max }} \right], 
\end{align}
where $\lambda$ is the Lagrange multiplier corresponding to constraint ${\rm{C_5}}$. The optimal solution $p_{n,k}^*\left( \tau \right)$ is given by
\begin{align}
\label{opp}
p_{n,k}^* \left( \tau \right)= \min \left[ {p_{n,k}^{\max },\frac{{{Q_n}\left( \tau \right){T_0}{W_k}\left( \tau \right)}}{{{\cal E}\left( \tau \right)\ln{2}}} - \frac{{{\sigma ^2}}}{{{h_{n,k}}\left( \tau \right)}}} \right].
\end{align} }

{We can notice that $p_{n,k}^*\left( \tau \right)$ is positively related to ${Q_n}\left( \tau \right)$, and is negatively related to ${h_{n,k}}\left( \tau \right)$ and ${\cal E}\left( \tau \right)$.}

{\subsubsection{Preference list construction} We define the preference of device $n$ towards channel $k$ as
\begin{align}
\label{flist}
{F_{n,k}}{|_{\varphi \left( n \right) = k}} = {f_{D_3}}\left[ {p_{n,k}^ * \left( \tau \right)\left| {{x_{n,k}}\left( \tau \right) = 1} \right.} \right] - {\Lambda _k},
\end{align}
where the virtual price ${\Lambda _k}$ reflects the matching cost of channel $k$. 
The preference list of device $n$, i.e., ${\cal {F}}_n$, is constructed by sorting all $K$ channels in descending order according to the preferences, i.e., ${F_{n,k}}{|_{\varphi \left( n \right) = k}}$, $\forall k \in {\cal K}$. The total set $\cal F$ is constructed as ${\cal F}=\left\{{\cal{F}}_n, \forall n \in {\cal N} \right\}$. }
{\subsubsection{Proposal and price rising} Denote $|\varphi(n)|$ as the size of $\varphi(n)$. If $\exists |\varphi(n)|<q$, the device $n \in {\cal N}$ will propose to the first $q-|\varphi(n)|$ channels in its preference list ${\cal{F}}_n$. Afterwards, if any channel $k\in{\cal K}$ receives only one proposal from a device, then they will be directly matched. Otherwise, if $k$ receives more than one proposal, add $k$ into set $\Omega$ and implement the price rising process to solve matching conflicts.}
{Each channel $k \in \Omega$ raises its price ${\Lambda _k}$ by $\Delta {\Lambda _k}$ to increase the matching cost, which is given by
\begin{align}
\label{pra}
{\Lambda _k}={\Lambda _k}+\Delta {\Lambda _k}.
\end{align}}
{Accordingly, all the devices proposed to $k$ update their preferences as (\ref{flist}) and renew their proposal strategies. Some devices proposed to channel $k$ may give it up due to the increasing matching cost. The price rising process will continue until only one device remains, which is eventually matched with channel $k$. Then, $k$ is removed from $\Omega$. If all the channels in ${\cal{F}}_n$ have been matched with other devices and are unavailable to $n$, then $\varphi(n)=\{n\}^{1\times q}$.}
{\subsubsection{Matching termination} The matching process will be finished until a stable matching is produced. The devices select the channels based on the derived $\varphi$.}
{\begin{algorithm}[htbp]
\caption{Joint Channel Selection and Power Allocation based on One-to-many Matching}
\label{matchinga}
\begin{algorithmic}[1]
\STATE \textbf{Input:} $N$, $K$, $T$, $M$, $q$.
\STATE \textbf{Output:} $\{x_{n,k}(t)\}$.
\STATE {\textbf{Phase 1:} Initialization }
\STATE {Set $\varphi = \emptyset$, $\Omega=\emptyset$, and ${\Lambda _k} = 0, \forall k \in {\cal K}$. }
\STATE {\textbf{Phase 2:} Power allocation}
\STATE {By temporarily matching every device $n$ with each channel, obtain the optimal power allocation result $\{p_{n,k}^* \left( \tau \right)\}$ as (\ref{opp}).}
\STATE {\textbf{Phase 3:} {Preference list construction }}
\STATE {Every device calculates its preference value toward each channel as (\ref{flist}).}
\STATE {The preference list of device $n$, i.e., ${\cal {F}}_n$, is constructed by sorting all $K$ channels in descending order according to the preferences, i.e., ${F_{n,k}}{|_{\varphi \left( n \right) = k}}$, $\forall k \in {\cal K}$.}
\STATE {\textbf{Phase 4:} {Proposal and price rising}}
\WHILE { $\exists |\varphi \left( n \right)| <q $}
\STATE {Each device $n \in {\cal N}$ proposes to the first $q-|\varphi \left( n \right)|$ channels in ${\cal{F}}_n$}.
\IF {any channel $k\in{\cal K}$ receives only one proposal from a device $n$,}
\STATE {the channel $k$ will be directly matched with device $n$.}
\ELSE 
\STATE {Add $k$ into set $\Omega$.}
\FOR {$k\in \Omega$}
\STATE { Each channel $k \in \Omega$ increases its price ${\Lambda _k}$ as (\ref{pra}).}
\STATE { All the devices proposed to $k$ update their preferences as (\ref{flist}) and renew their proposal strategies.}
\ENDFOR
\ENDIF 
\ENDWHILE
\STATE {\textbf{Phase 5:} {Matching termination}} 
\STATE The matching process will be finished until a stable matching is produced. 
\end{algorithmic}
\end{algorithm}}

%%%%%%%%%%%%%%%%%%%%%%%%%%%%%%%%%%%%%%%%%%%%%%%%%%%%%%%%%%%%%%%%%%%%%%%%%%%%%%%%%%%%%%%%%%%%%%%%%%%%%%%%%%%%%%%
\section{Performance analysis}
\label{sec:6}
In this section, some theoretical properties in terms of optimality performance, convergence performance, and computational complexity are analyzed.
\subsection{Tradeoff between Queue Stability and Utility Maximization}
\label{sec:61}

\begin{theorem}\label{Theorem2}
Algorithm \ref{Lyapunov optimization} achieves a $\left[ {O\left( V \right),O\left( {{\raise0.7ex\hbox{$1$} \!\mathord{\left/{\vphantom {1 V}}\right.\kern-\nulldelimiterspace} \!\lower0.7ex\hbox{$V$}}} \right)} \right]$ tradeoff between queue stability and utility maximization by adjusting the control parameter $V$. The time-average data queue backlog, time-average energy queue backlog, and time-average network utility are bounded by
\begin{align}
&\mathop {\lim }\limits_{M \to \infty } \displaystyle\frac{1}{{MT}}\mathbb{E}\left[ {{\displaystyle\sum\limits_{m = 1}^M {\displaystyle\sum\limits_{\tau  = (m-1)T+1}^{mT} }{\displaystyle\sum\limits_{n = 1}^N {{Q_n}\left( \tau  \right)} } } } \right] \nonumber \\
& \le \displaystyle\frac{B}{{2{\delta _1}}} + \displaystyle\frac{{V\left( {{f_{\max }} - {f_{opt}}} \right)}}{{{\delta _1}}},\\
&\mathop{\lim }\limits_{M\to\infty }\displaystyle\frac{1}{{MT}}\mathbb{E}\left[\displaystyle\sum\limits_{m = 1}^M {\displaystyle\sum\limits_{\tau  = (m-1)T+1}^{mT} {{E}\left( \tau\right)}}\right] \nonumber \\
&\ge  {E_{\max}}-\displaystyle\frac{B}{{2{\delta _2}}} -\frac{{V\left({{f_{\max }}-{f_{opt}}}\right)}}{{{\delta _2}}},\\
&\mathop {\lim }\limits_{M \to \infty } \frac{1}{{MT}}\mathbb{E}\left[ {\displaystyle\sum\limits_{m = 1}^M {\displaystyle\sum\limits_{\tau  = (m-1)T+1}^{mT}{f\left( \tau  \right)}}} \right] \ge {f_{opt}} - \frac{B}{{2V}},
\end{align}
where ${f_{\max }}$ is the finite constant to bound $\mathbb{E}\left[f(\tau)\right]$, and $f_{opt}$ is the theoretical optimum of \bf{P1} \cite{Guo2017Quality}.
\end{theorem}
\begin{IEEEproof}
{See Appendix \ref{APPB}.}
\end{IEEEproof}

\begin{theorem}\label{Theorem3}
The joint channel selection and power allocation algorithm produces a stable matching between devices and channels within finite iterations.
\end{theorem}
\begin{IEEEproof}
{See Appendix \ref{APPC}.}
\end{IEEEproof}

\subsection{Convergence of ADMM-based Low-complexity Rate Control Algorithm}

\begin{theorem}\label{Theorem4}
The residual convergence, objective convergence, and dual variable convergence are expressed as follows:
\begin{enumerate}[1)]
\item Residual convergence: The primal and dual residuals converge to 0 as $i \to \infty$, which implies that the iterations approach feasibility.
\item Objective convergence: The objective function of $\mathbf{P7}$ eventually converges to the primal optimal value under the stopping criterion as $i \to \infty$.
\item Dual variable convergence: The dual variable ${{y}}^{i+1}$ eventually  converges to the dual optimal value under the stopping criterion as $i \to \infty$.
\end{enumerate}
\end{theorem}
\begin{IEEEproof}{See Appendix \ref{APPD}.}
\end{IEEEproof}

\subsection{Computational Complexity}
\subsubsection{Computational complexity of energy management}
The computational complexity of linear programming is in the linear order with the number of optimization variables. Similarly, the energy management problem is also a linear programming problem with two variables optimized over a total of $M$ energy frames. Thus, its computational complexity is $O(2M)$.
\subsubsection{Computational complexity of rate control}
Rate control is optimized at each data slot with $N$ optimization variables. Thus, updating primal and dual variables introduces a complexity of $O\left[\max \left( {{l_r},N - {l_r}} \right)\right]$. Assuming that ${\bf{x}}_r$, ${\bf{z}}_r$, and ${{\mu}}$ are updated $\xi$ times before reaching convergence, the total complexity of rate control is $O\left[ {\max \left( {{l_r},N - {l_r}} \right) \times {MT}{\xi}} \right]$.
\subsubsection{Computational complexity of joint channel selection and power allocation}
The complexities for each device to acquire the preferences and construct the preference list are $O\left( K\right)$ and $O\left(K \log {\left( K \right)} \right)$, respectively. Assuming that the number of iterations required for resolving the conflict in the price rising process is $\varsigma$, and there are $\max\left({N,K}\right)$ conflict elements in the price rising process, the complexityof joint channel selection and power allocation is $O\left\{ {MT\left\{ {\max (N,K) \times \varsigma  + \left[ {NK + NK\log \left( K \right)} \right]} \right\}} \right\}$.

\begin{table}[ht]
\caption{Simulation Parameters.}
\label{tab:1}
\begin{center}
\begin{tabular}{|p{120pt}|p{100pt}|}
\hline
\textbf{Parameter}&\textbf{Value}\\
\hline
Number of devices &$N=5$ \\
\hline
Number of channels &$K=12$ \\
\hline
Channel bandwidth &$W=1$ MHz \\
\hline
Tunable weight of DMU &$V=100$ \\
\hline
Number of energy frames &$M=200$ \\
\hline
Number of data slots at each energy frame&$T=5$ \\
\hline
One data slot duration &$T_0=1$ second \\
\hline
Upper bound of purchased energy &$g_{\max}=2.5$  \textbf{J}\\
\hline
Capacity of recharge battery &${E _{\max }} = 5$ J\\
\hline
Maximum sum of arrival data &${R _{\max }} = 20$ Mbps\\
\hline
Service weight parameter &$\chi  = \left[ {0.1,0.15,0.2,0.25,0.3} \right]$\\
\hline
Penalty factor of utility &$\beta  = \left\{0,5000\right\}$ \\
\hline
The initial state of data queue &${Q_n}\left( 1 \right) = 3$ Mbits \\
\hline
The initial state of energy queue &${E}\left( 1 \right) = 2$ J \\
\hline
Time-average delay constraint &$d_n^* = 10$ microseconds \\
\hline
Quota restriction &${\bf{q}} = \left[ {3,1}\right]$  \\
\hline
\end{tabular}
\end{center}
\end{table}

\section{Simulation Results}
\label{sec:7}
In this section, we verify the proposed two-timescale resource allocation algorithm through simulations. Simulation parameters are summarized in Table \ref{tab:1}\cite{Shan2016Energy, Guo2017Quality}. Based on \cite{Cui2015Grid}, the energy frame duration is about five to ten times of the data slot duration. Hence, $T$ is set as five times of the data slot duration. The proposed algorithm can be adaptable to different values of $T$ including scenario where $T\gg 1$. The data storage capacity of the BS is set as $500$ GB. In simulations, we set the price of harvested energy as $0$, i.e., $\kappa\left(\tau \right)=0$. This is reasonable since the energy is harvested from the external renewable energy sources rather than purchased from the power grid. The BS does not need to pay any money to the grid company. Similar assumption is also adopted in \cite{ZhangDistributed}. Moreover, we also consider the scenario where $\kappa(\tau)\neq 0$. Based on (\ref{eq20}) and (\ref{eq21}), the proposed algorithm can be extended to more general price settings. We consider different types of electricity price. Three heuristic algorithms are used as baselines for comparison purpose. In the baseline $1$ algorithm, the optimization of channel selection is neglected, and channels are allocated to devices randomly \cite{Zhou2018Social}. The baseline $2$ algorithm only maximizes the time-average QoE of network, while the minimization of energy cost is neglected, i.e., $\beta=0$. In the baseline $3$ algorithm, the rate control problem is solved by the convex optimization toolbox \cite{Guo2017Quality}, i.e., the CVX toolbox.

\begin{figure*}[t]
\centering
\begin{minipage}[t]{0.32\textwidth}
\centering
\includegraphics[width=1\linewidth]{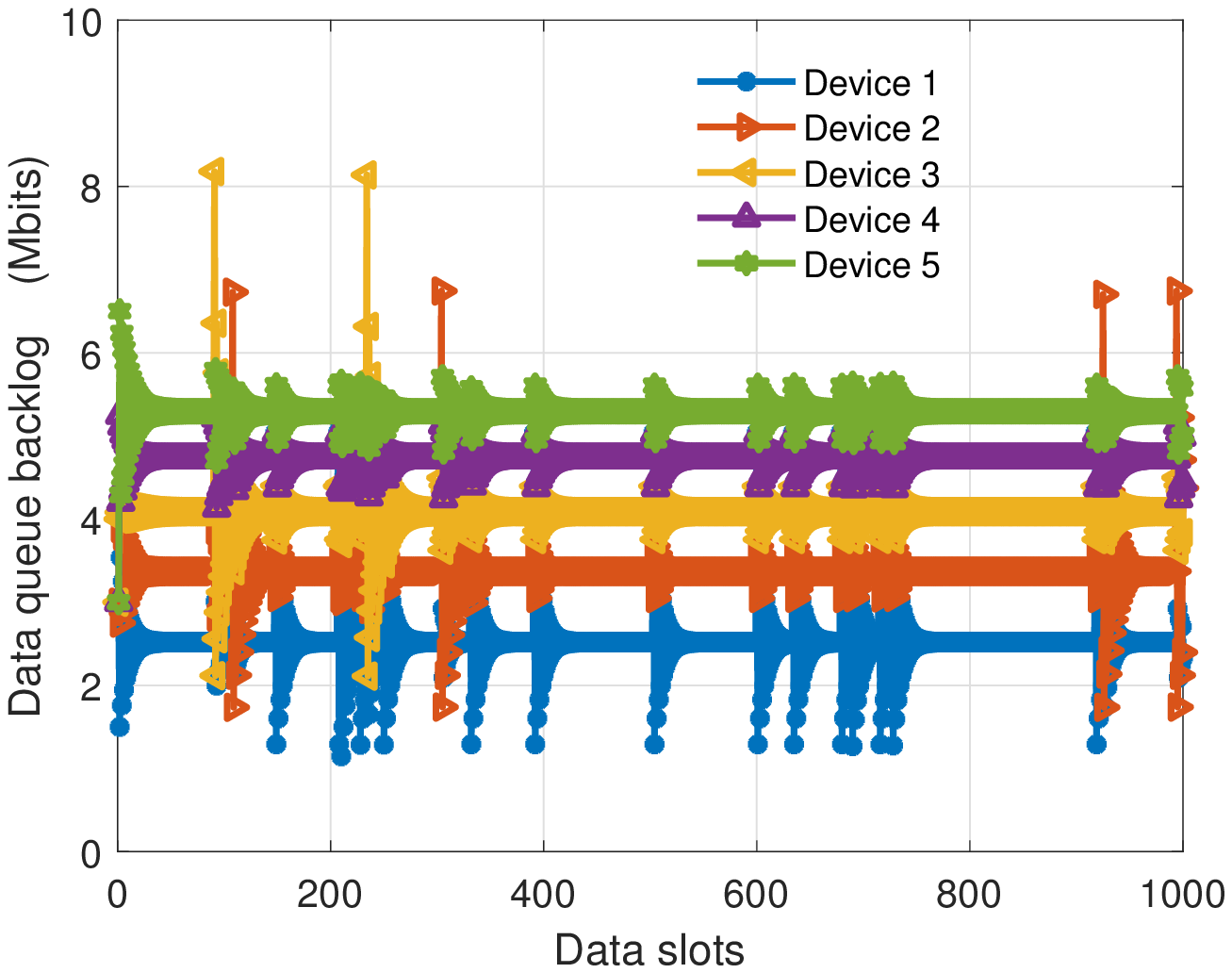}
\caption{Data queue backlog of the proposed algorithm.}
\label{fig3:side:3}
\end{minipage}
\begin{minipage}[t]{0.32\textwidth}
\centering
\includegraphics[width=1\linewidth]{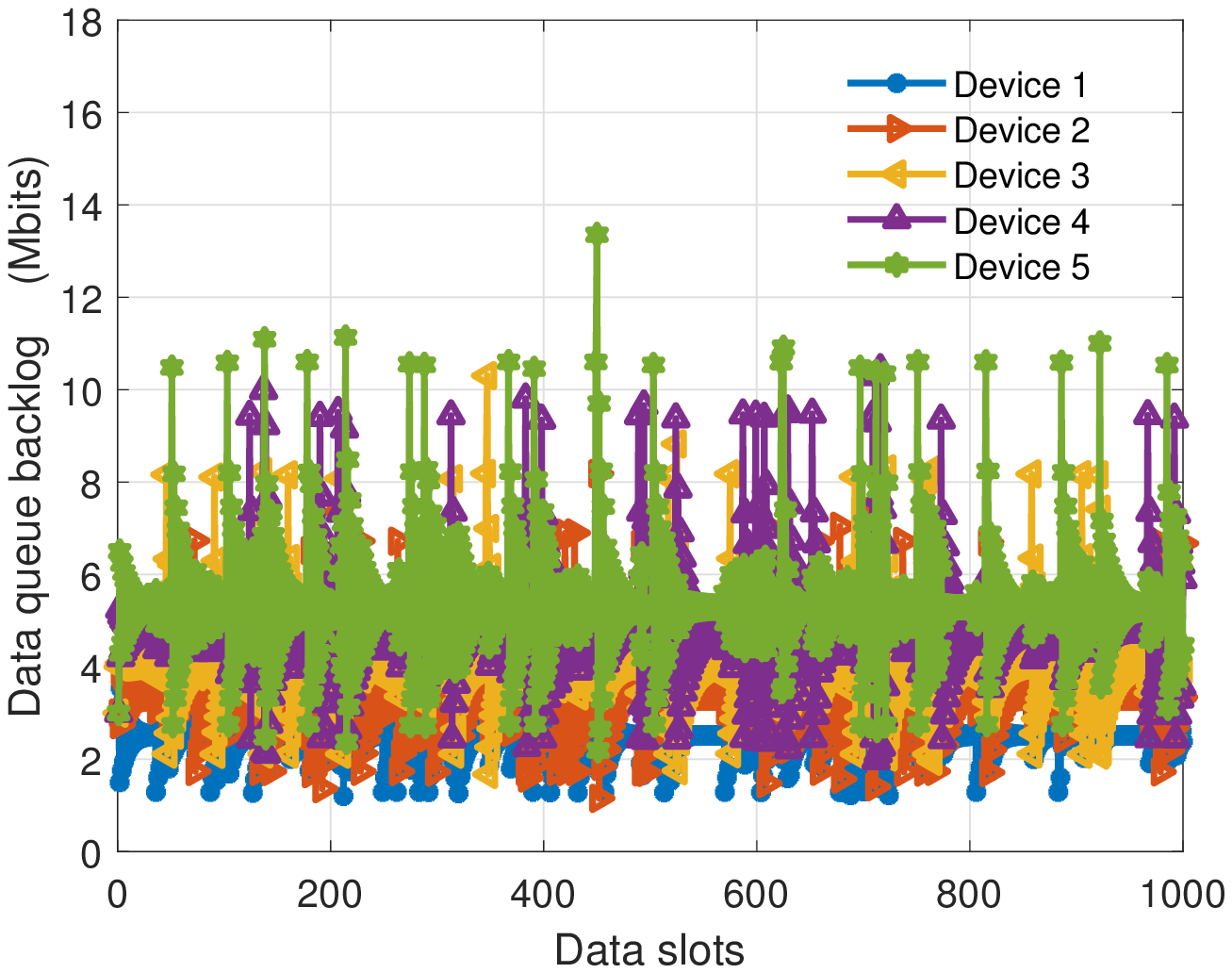}
\caption{Data queue backlog of the baseline $1$ algorithm.}
\label{fig3:side:4}
\end{minipage}
\begin{minipage}[t]{0.32\textwidth}
\centering
\includegraphics[width=1\linewidth]{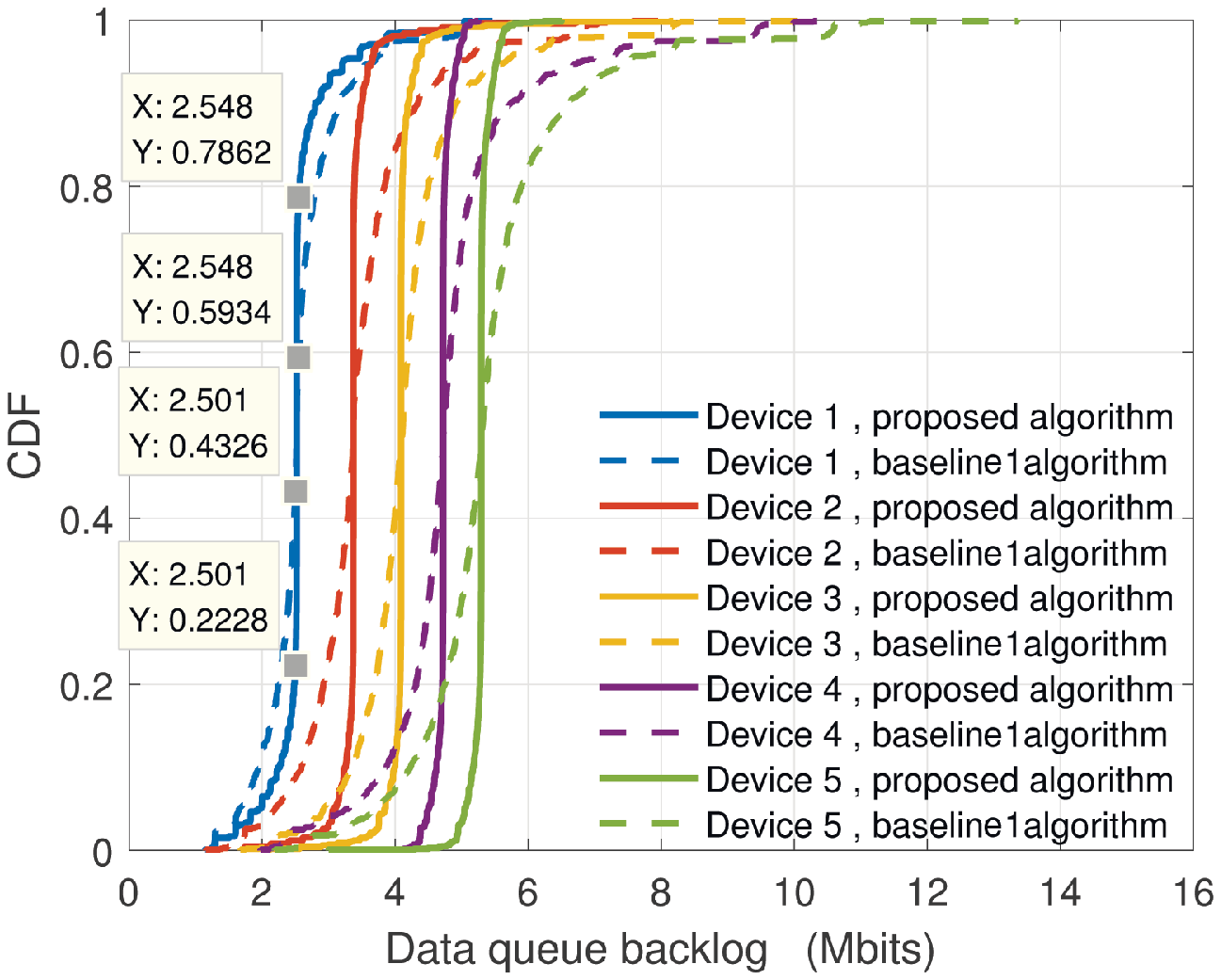}
\caption{CDF of data queue backlog.}
\label{fig3:side:5}
\end{minipage}%
\end{figure*}

\begin{figure*}[t]
\centering
\begin{minipage}[t]{0.32\textwidth}
\centering
\includegraphics[width=1\linewidth]{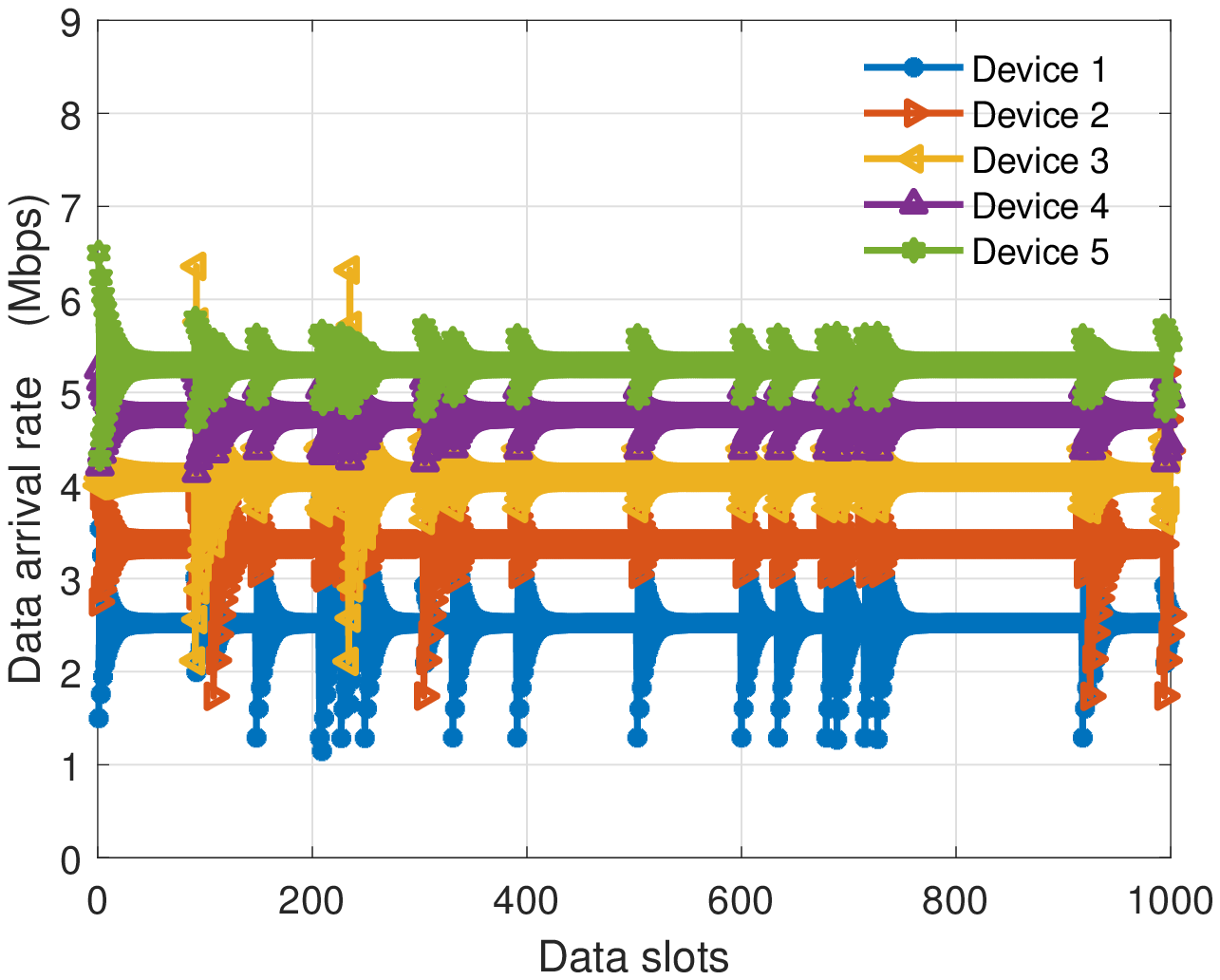}
\caption{Data arrival rate of the proposed algorithm.}
\label{fig4:side:6}
\end{minipage}
\begin{minipage}[t]{0.32\textwidth}
\centering
\includegraphics[width=1\linewidth]{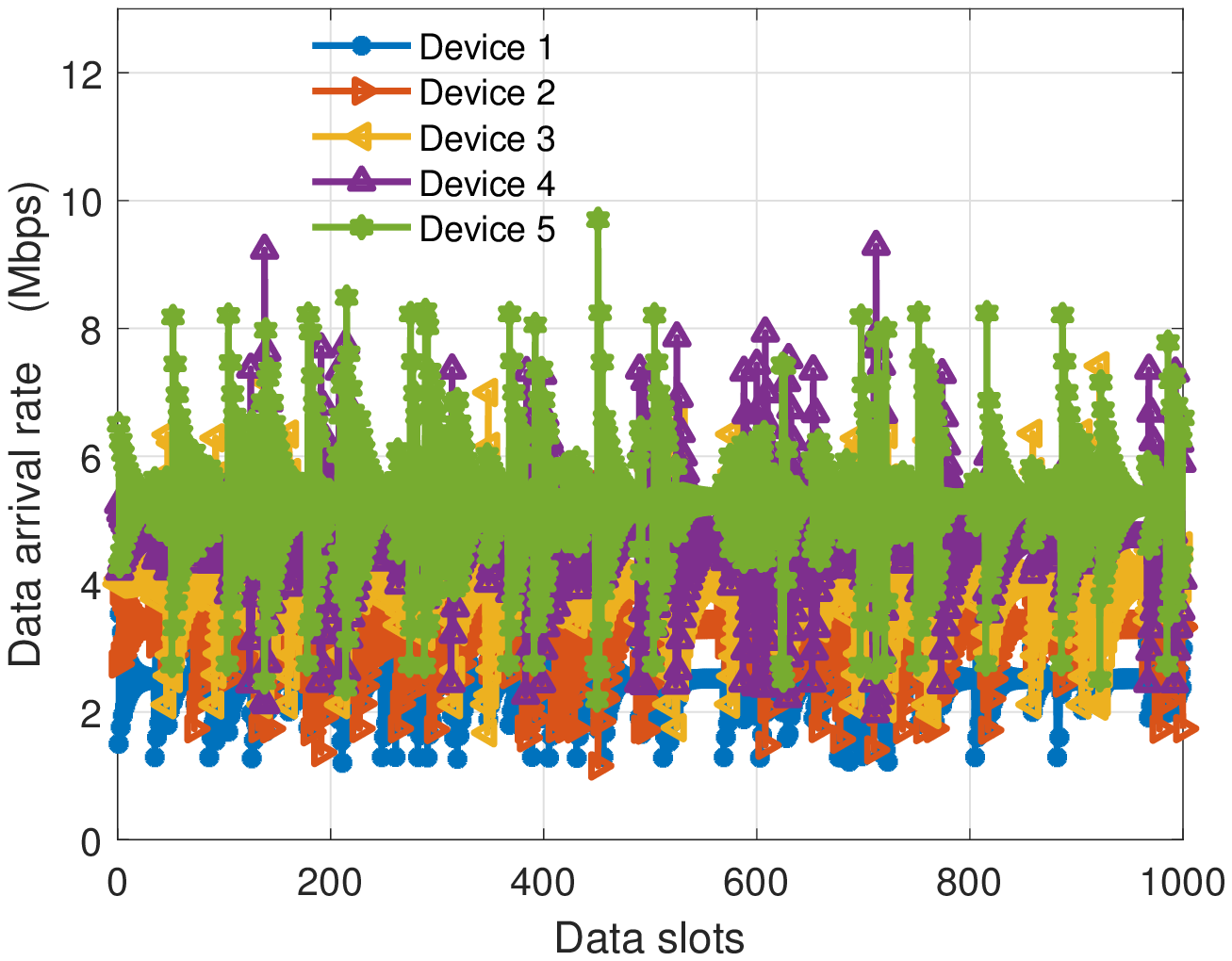}
\caption{Data arrival rate of the baseline $1$ algorithm.}
\label{fig4:side:7}
\end{minipage}
\begin{minipage}[t]{0.32\textwidth}
\centering
\includegraphics[width=1\linewidth]{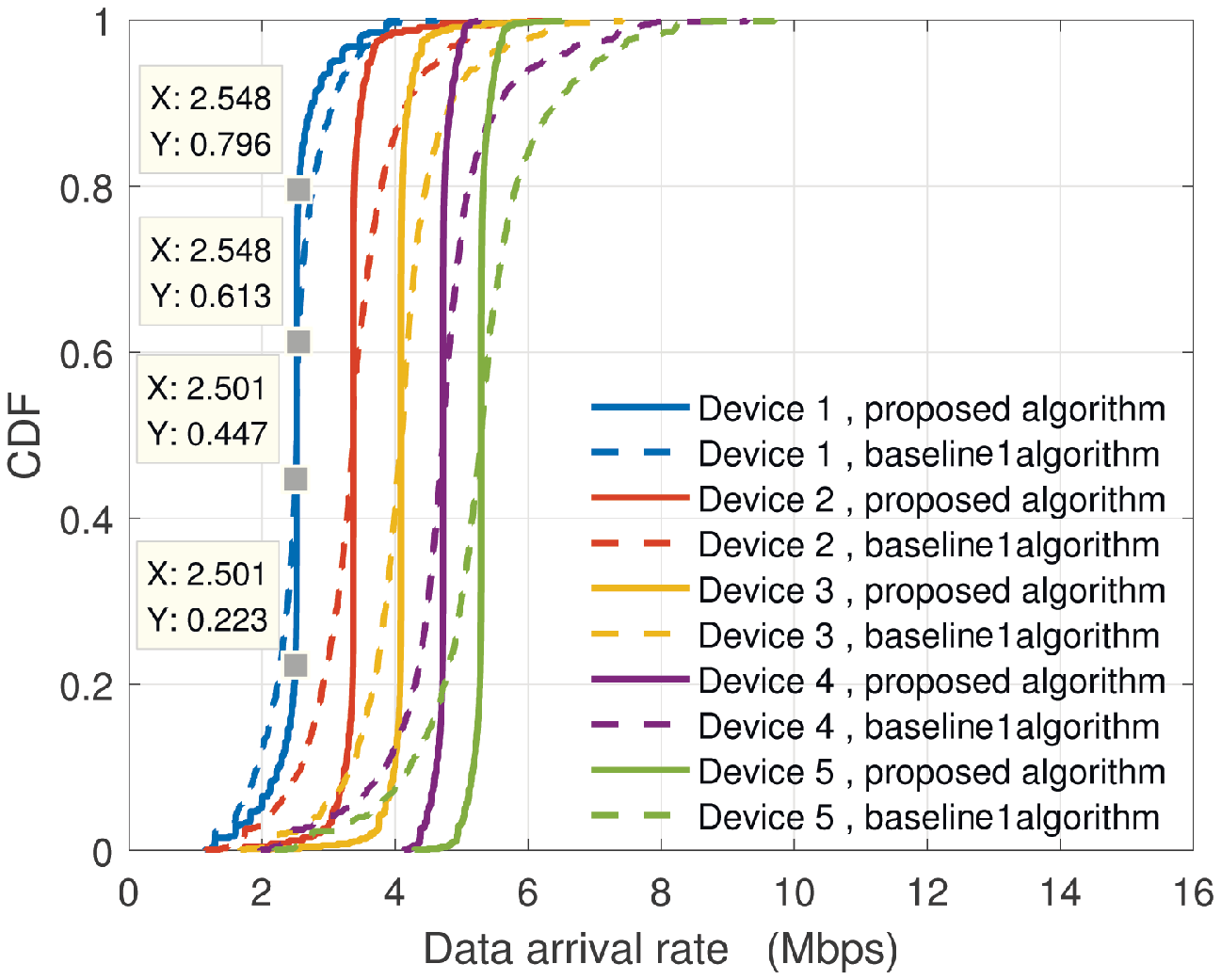}
\caption{CDF of data arrival rate.}
\label{fig4:side:8}
\end{minipage}%
\end{figure*}

\begin{figure*}[t]
\centering
\begin{minipage}[t]{0.32\textwidth}
\centering
\includegraphics[width=1\linewidth]{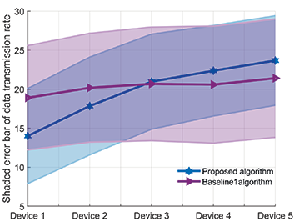}
\caption{Shaded error bar of transmission rate.}
\label{fig5:side:11}
\end{minipage}%
\begin{minipage}[t]{0.32\textwidth}
\centering
\includegraphics[width=1\linewidth]{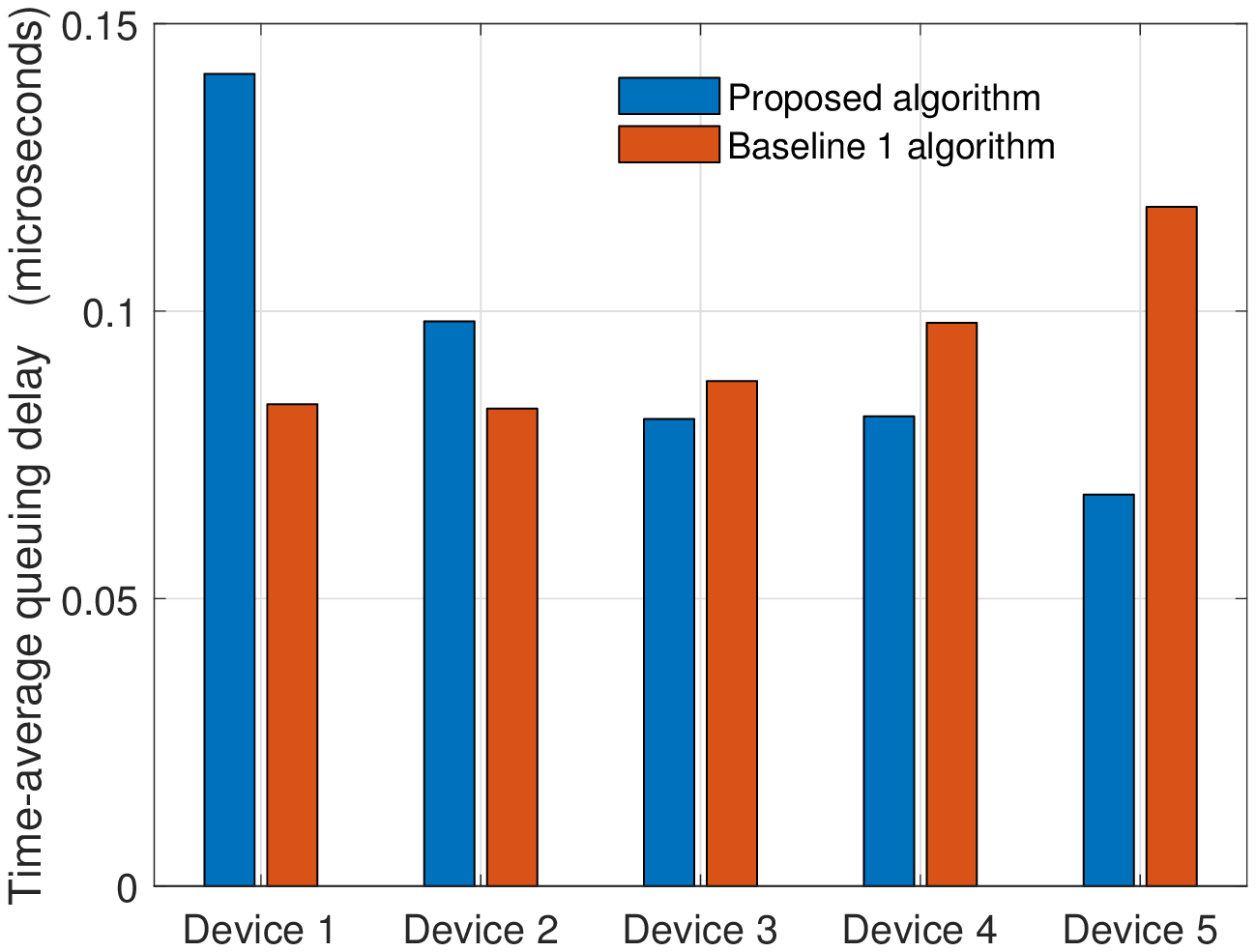}
\caption{Time-average queueing delay.}
\label{fig_delay_BAR}
\end{minipage}%
\begin{minipage}[t]{0.32\textwidth}
\centering
\includegraphics[width=1\linewidth]{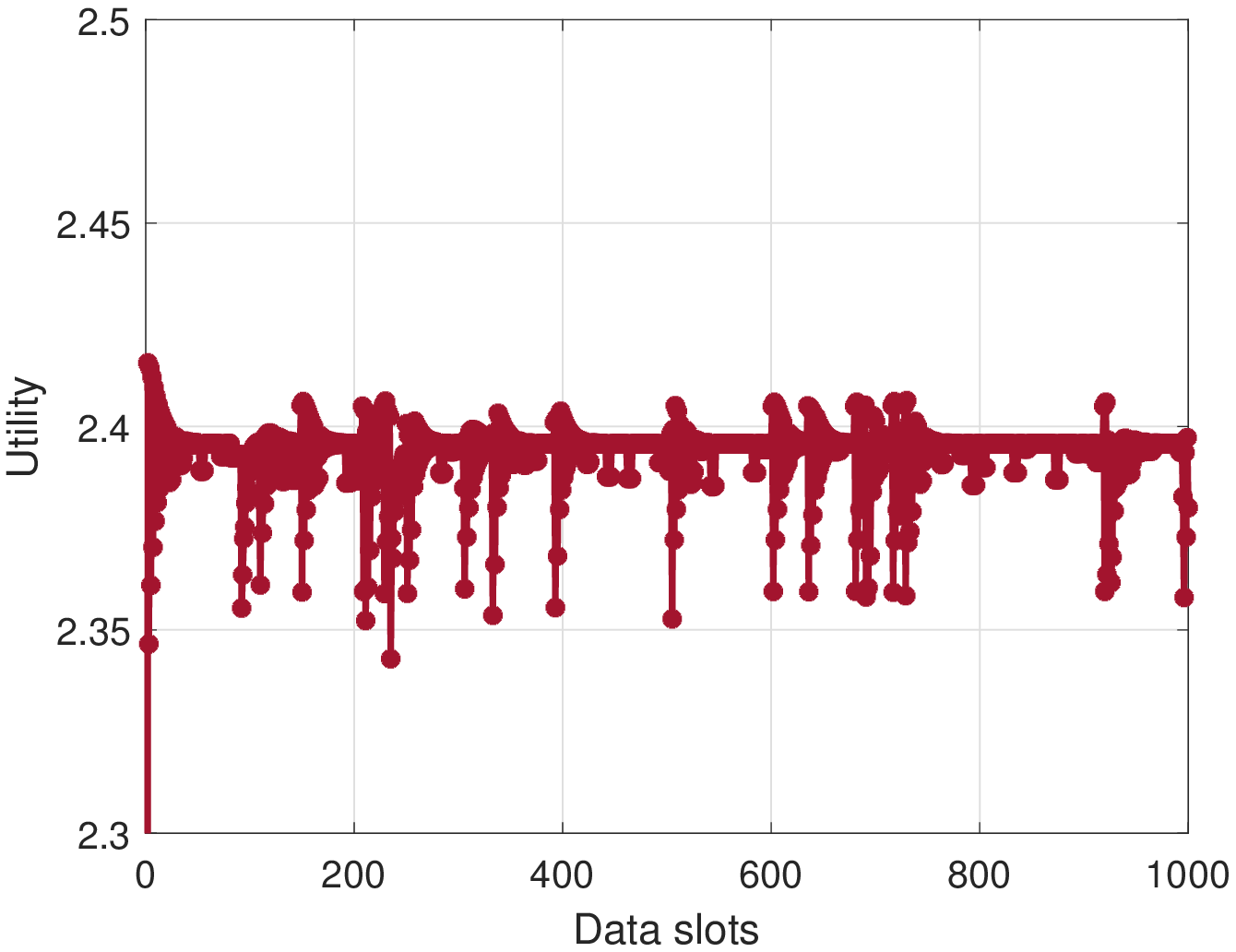}
\caption{Network utility of the proposed algorithm.}
\label{fig11:side:24}
\end{minipage}%
\end{figure*}

\begin{figure*}[t]
\centering
\begin{minipage}[t]{0.32\textwidth}
\centering
\includegraphics[width=1\linewidth]{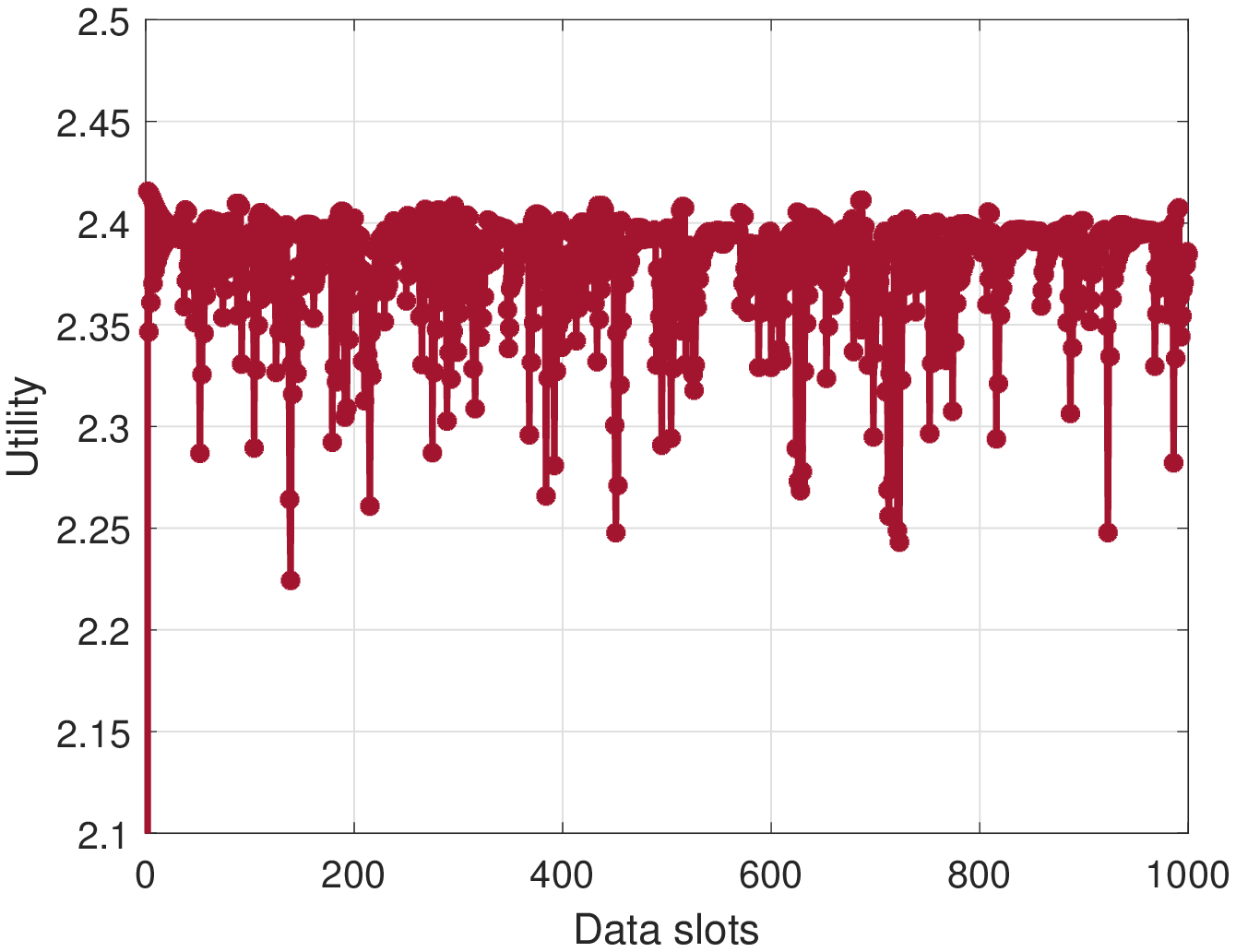}
\caption{Network utility of the baseline $1$ algorithm.}
\label{fig11:side:25}
\end{minipage}
\begin{minipage}[t]{0.32\textwidth}
\centering
\includegraphics[width=1\linewidth]{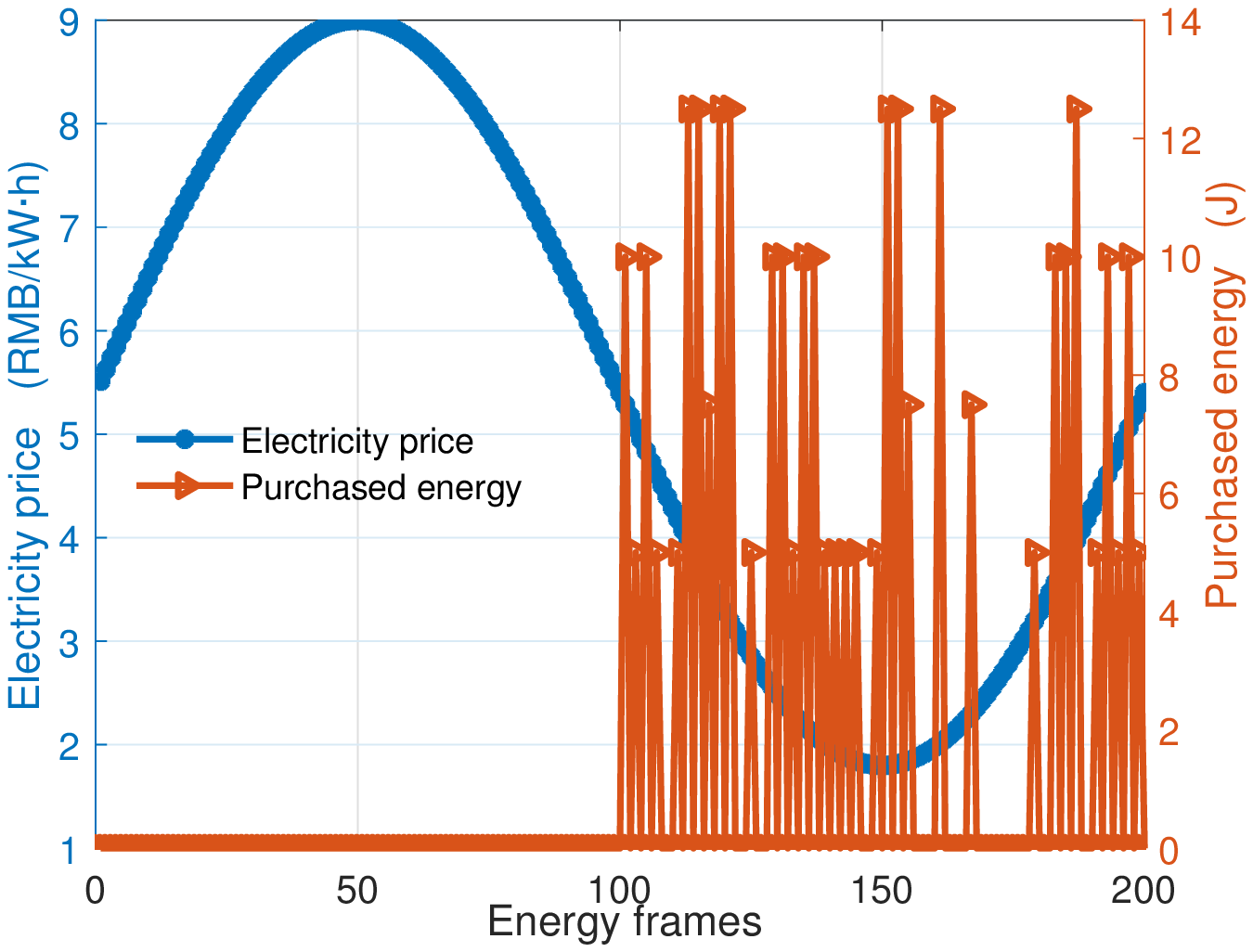}
\caption{Purchased energy of the proposed algorithm under sinusoid-based electricity price model.}
\label{fig_BETA5000_adapt}
\end{minipage}
\begin{minipage}[t]{0.32\textwidth}
\centering
\includegraphics[width=1\linewidth]{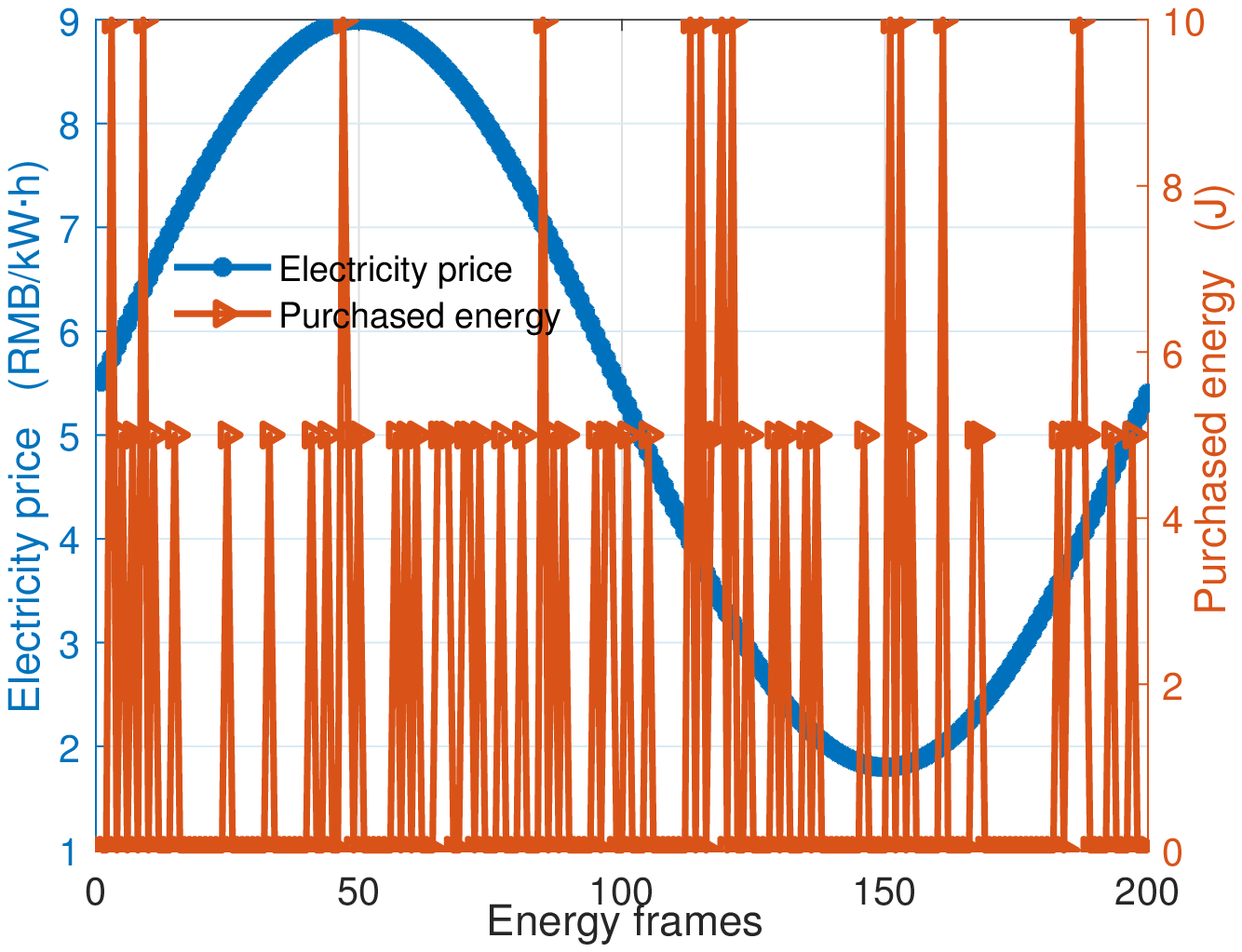}
\caption{Purchased energy of the baseline $2$ algorithm under sinusoid-based electricity price model.}
\label{fig_BETA0_adapt}
\end{minipage}
\end{figure*}

%%%%
\begin{figure*}[t]
\centering
\begin{minipage}[t]{0.32\textwidth}
\centering
\includegraphics[width=1\linewidth]{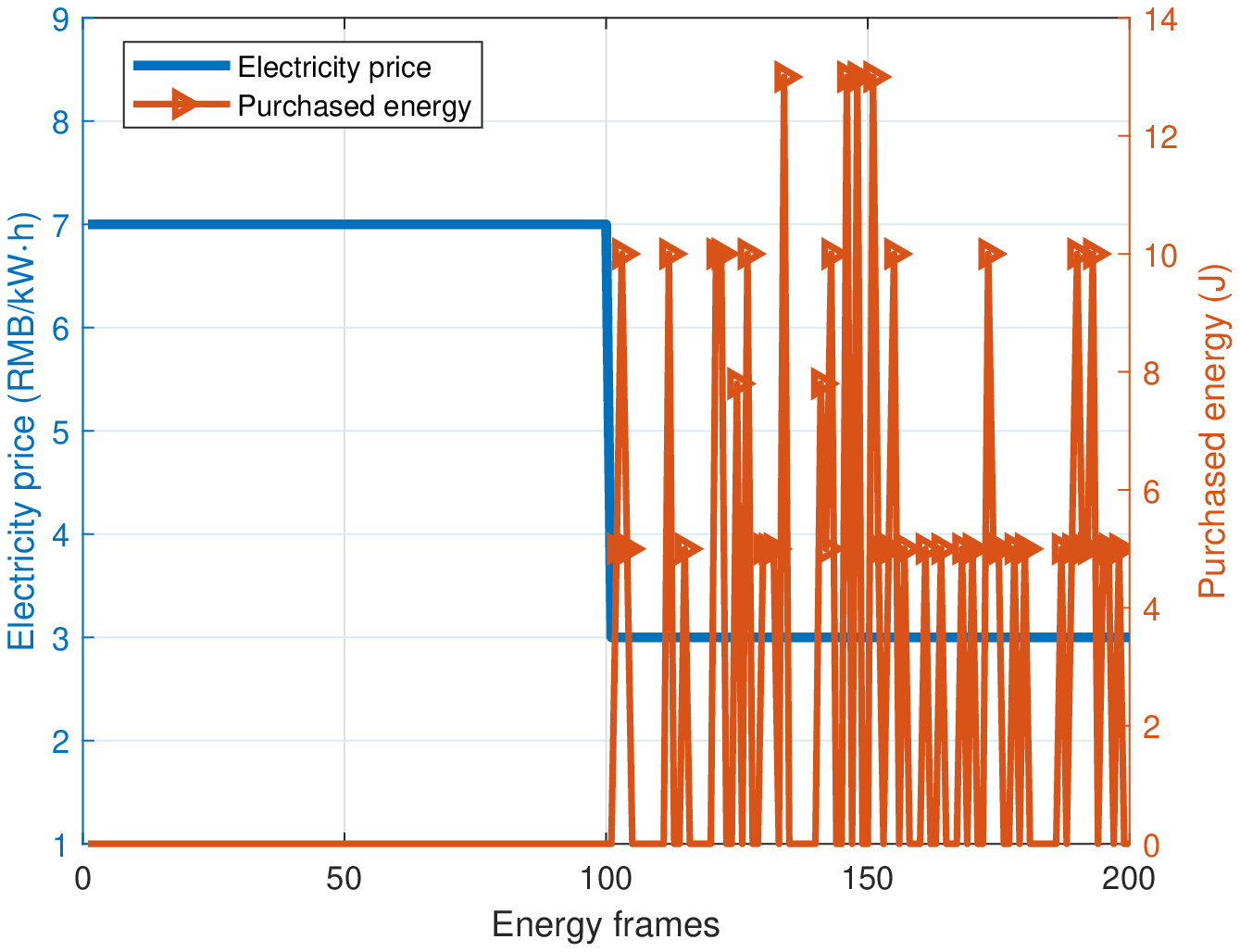}
\caption{{Purchased energy of the proposed algorithm under two-tier electricity price model.}}
\label{twoprice1}
\end{minipage}
\begin{minipage}[t]{0.32\textwidth}
\centering
\includegraphics[width=1\linewidth]{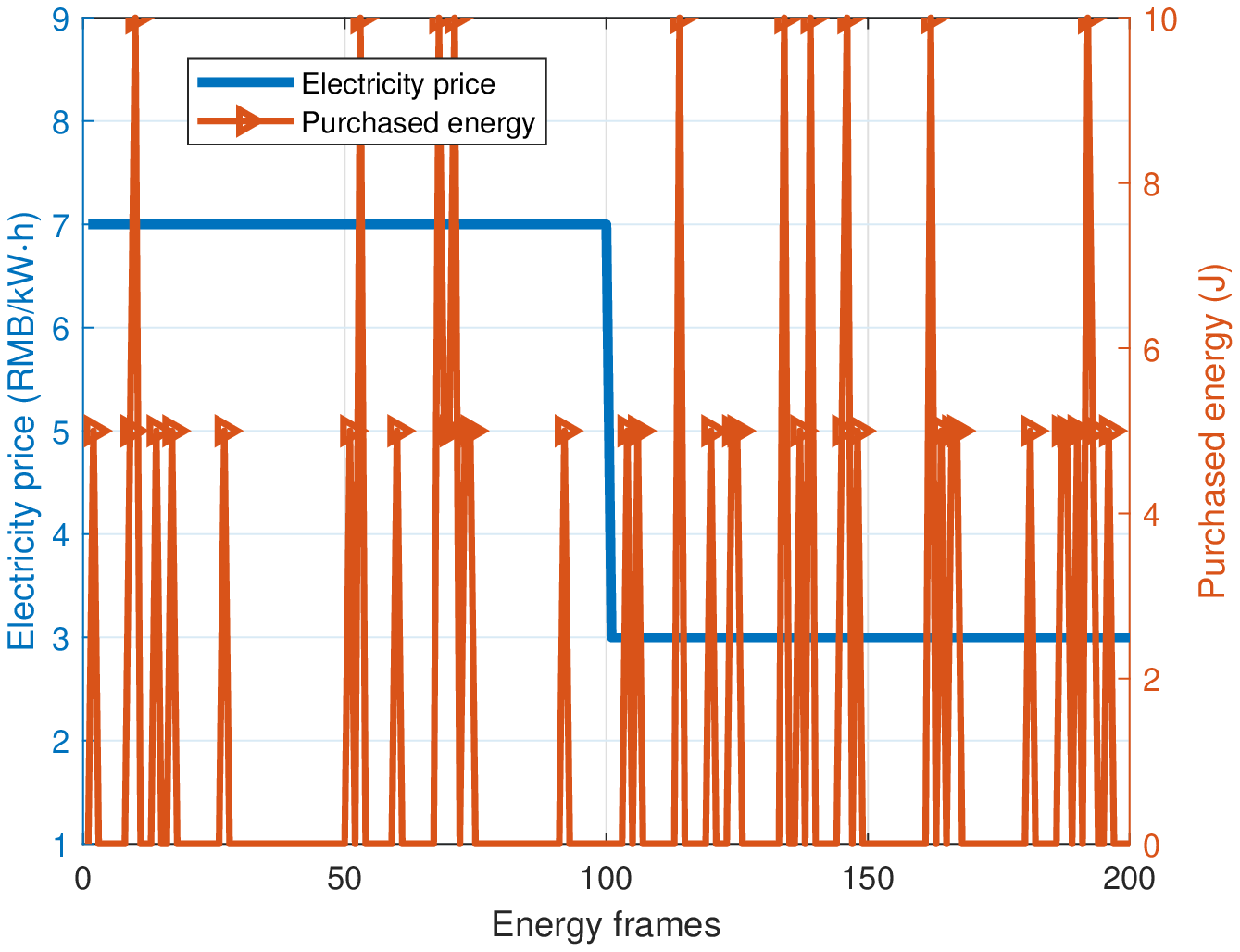}
\caption{{Purchased energy of the baseline $2$ algorithm under two-tier electricity price model.}}
\label{twoprice2}
\end{minipage}
\begin{minipage}[t]{0.32\textwidth}
\centering
\includegraphics[width=1\linewidth]{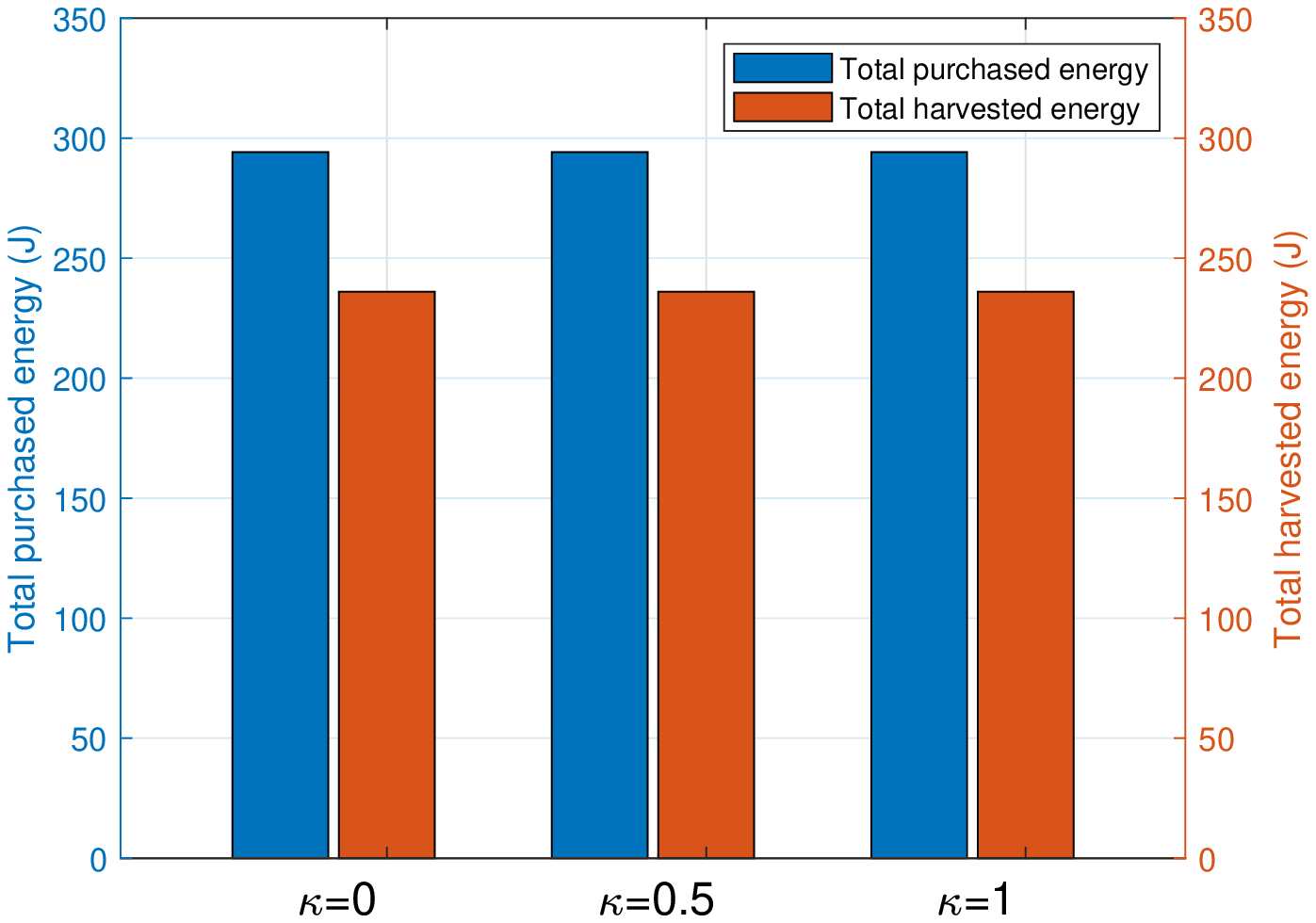}
\caption{{Total purchased energy and total harvested energy versus $\kappa$.}}
\label{kappa_price}
\end{minipage}
\end{figure*}

\begin{figure*}[t]
\centering
\begin{minipage}[t]{0.32\textwidth}
\centering
\includegraphics[width=1\linewidth]{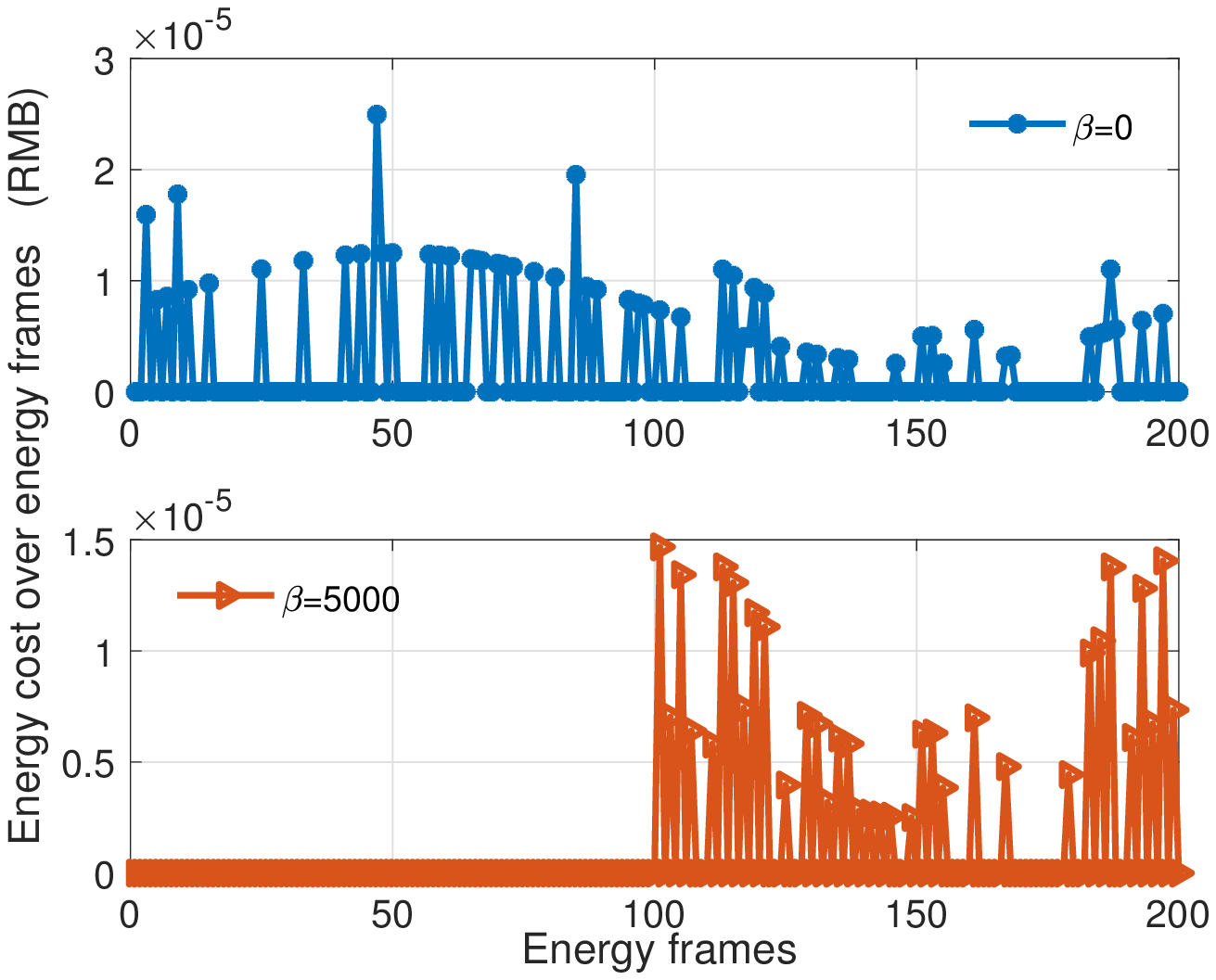}
\caption{Energy cost of the proposed algorithm and the baseline $2$ algorithm.}
\label{fig_Energycost}
\end{minipage}
\begin{minipage}[t]{0.32\textwidth}
\centering
\includegraphics[width=1\linewidth]{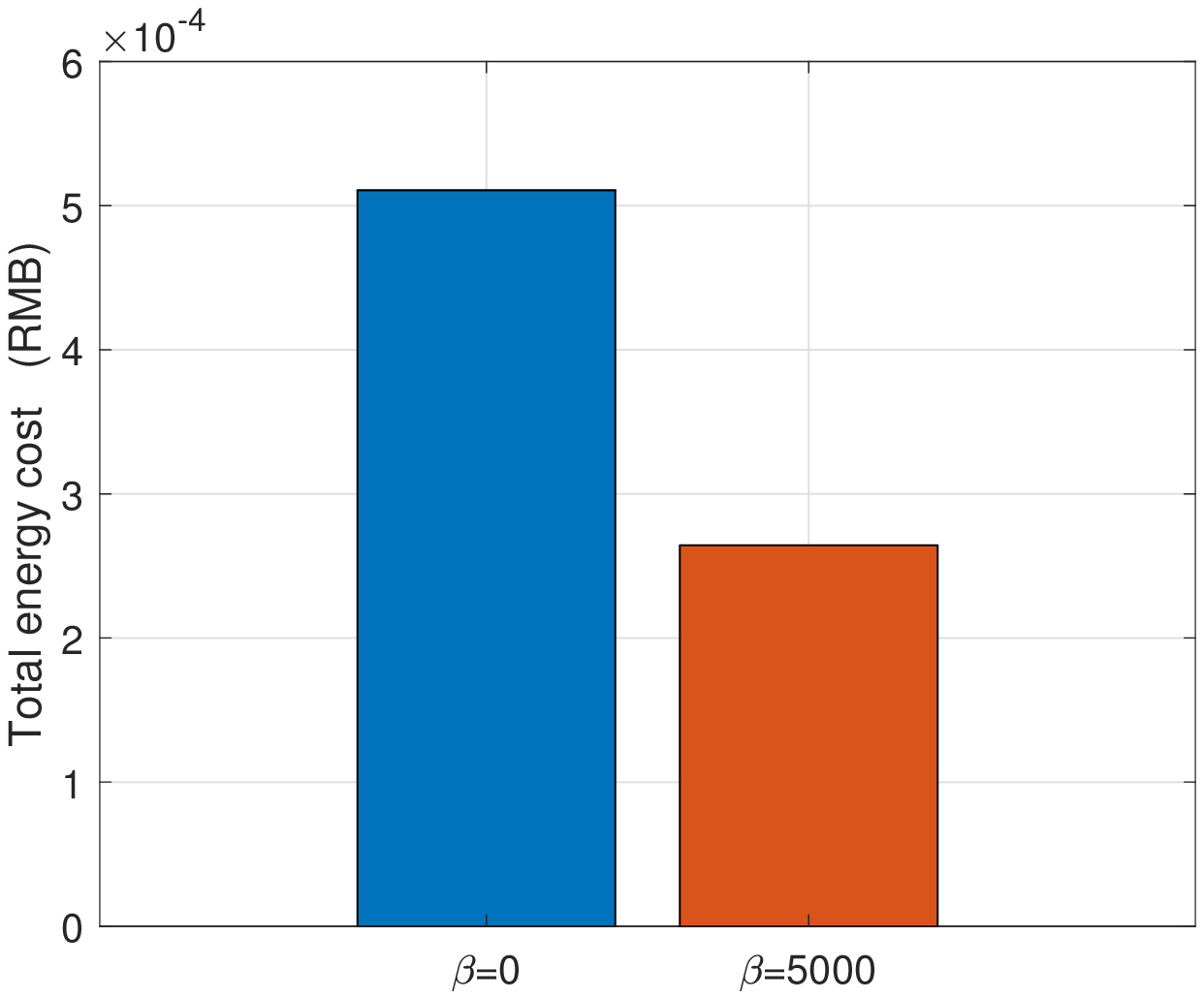}
\caption{Total energy cost over  $200$ ${\rm{ }}$ energy frames.}
\label{fig_Energycost_BAR}
\end{minipage}%
\begin{minipage}[t]{0.32\textwidth}
\centering
\includegraphics[width=1\linewidth]{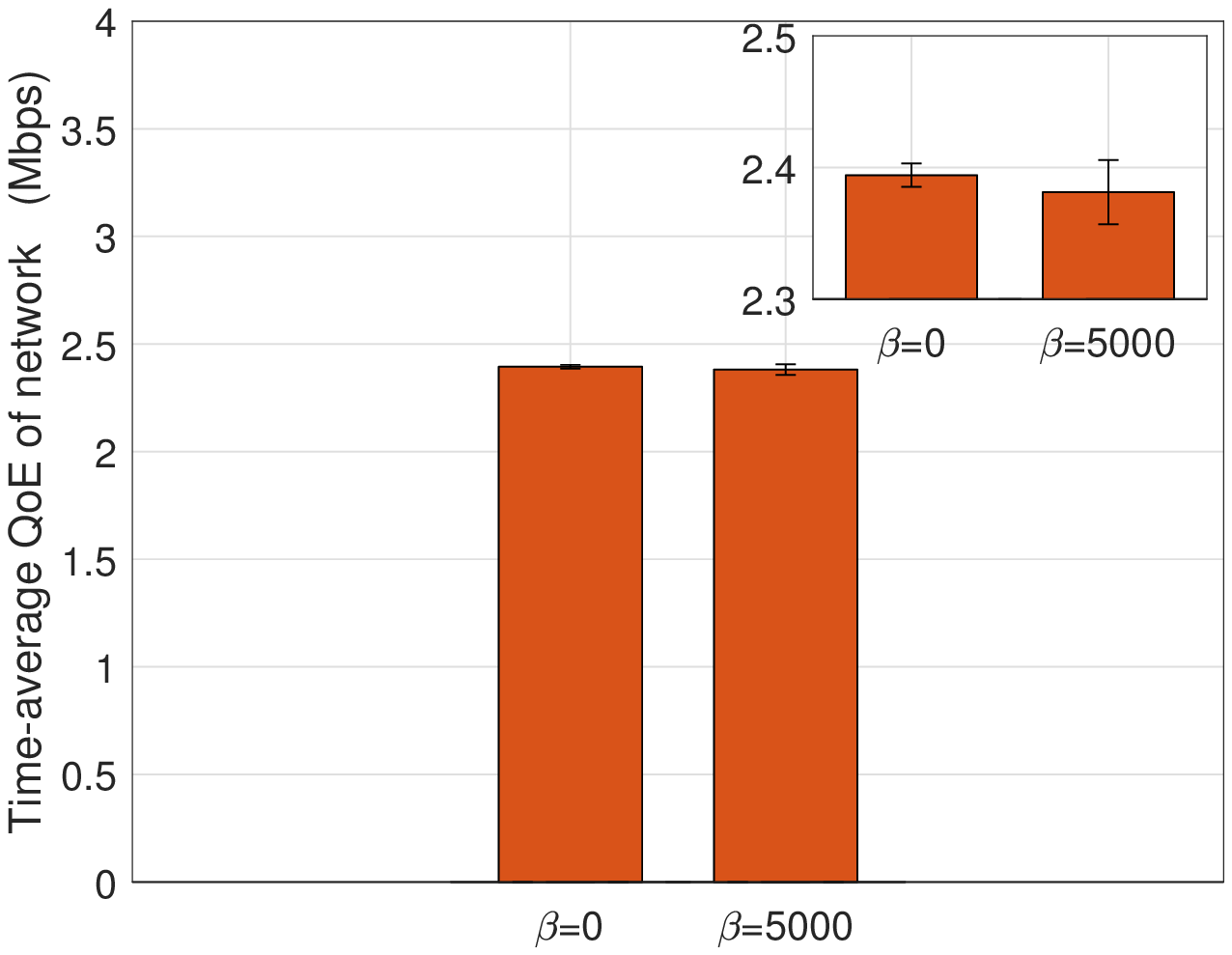}
\caption{Time-average QoE of network with standard deviation.}
\label{fig_QOE_BAR}
\end{minipage}
\end{figure*}

\begin{figure*}[t]
\centering
\begin{minipage}[t]{0.32\textwidth}
\centering
\includegraphics[width=1\linewidth]{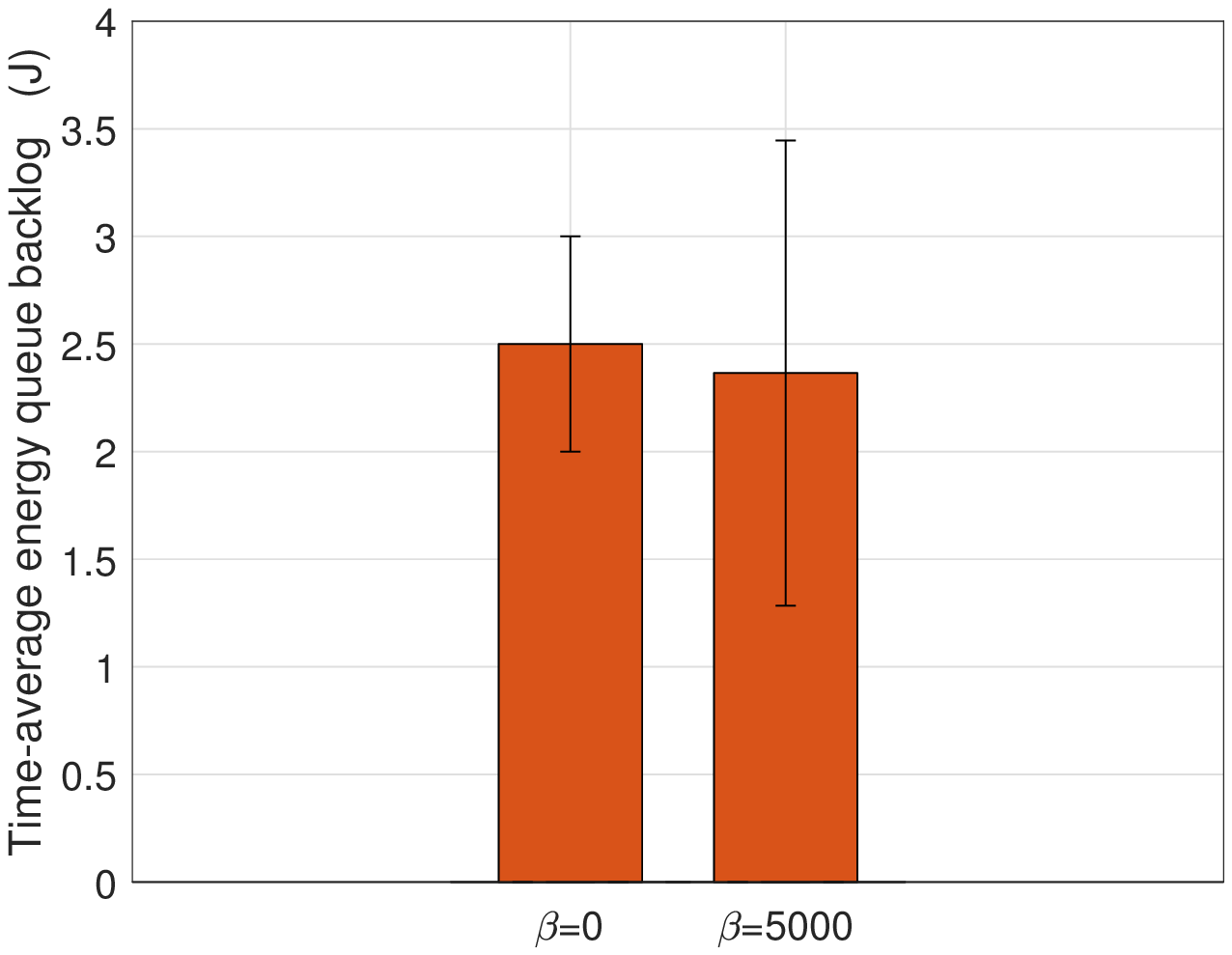}
\caption{Time-average energy queue backlog with standard deviation.}
\label{fig_EQEQ_BAR}
\end{minipage}
\begin{minipage}[t]{0.32\textwidth}
\centering
\includegraphics[width=1\linewidth]{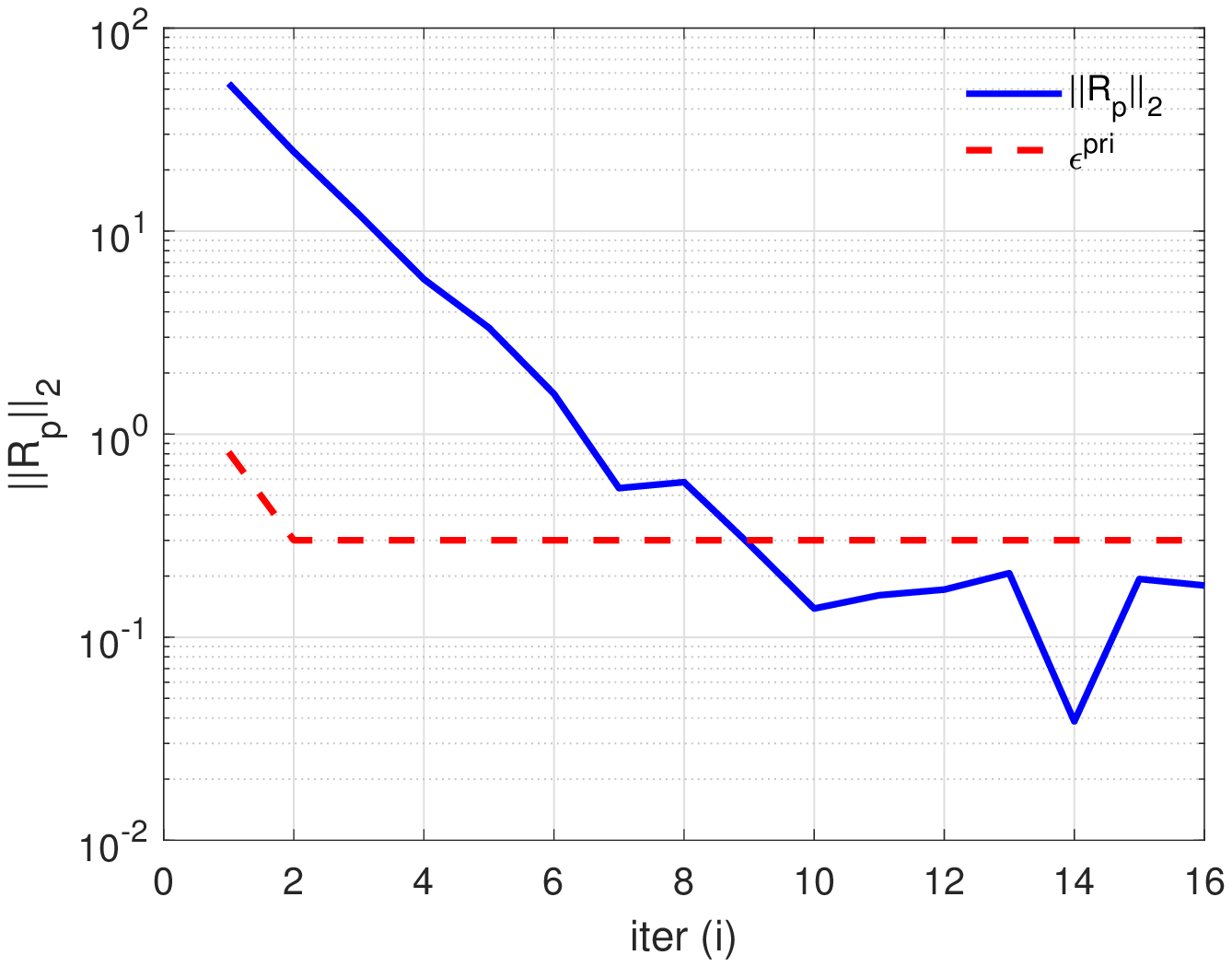}
\caption{Convergence of primal variables.}
\label{fig_admmRP}
\end{minipage}
\begin{minipage}[t]{0.32\textwidth}
\centering
\includegraphics[width=1\linewidth]{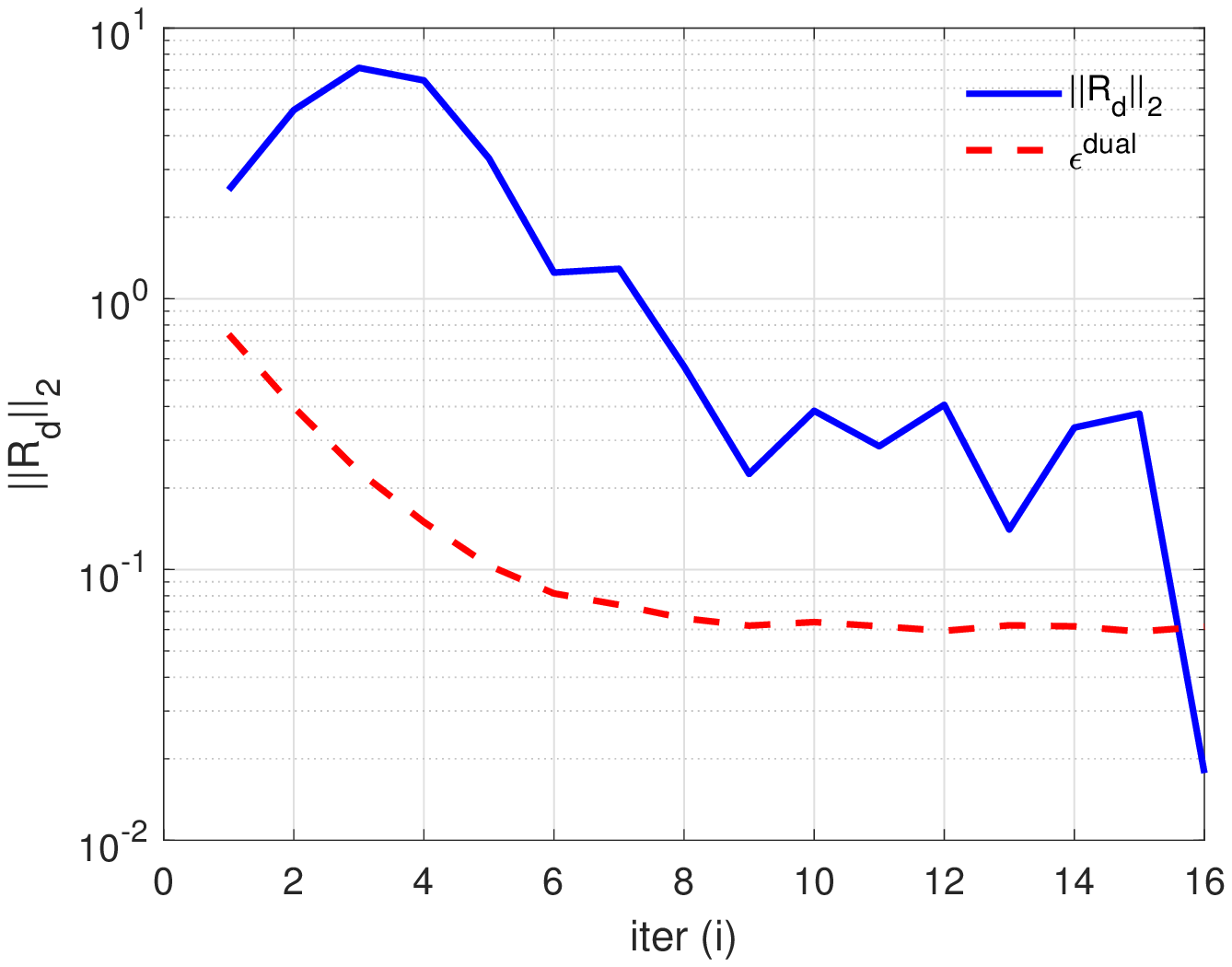}
\caption{Convergence of dual variables.}
\label{fig_admmRD}
\end{minipage}
\end{figure*}

\begin{figure*}[t]
\centering
\begin{minipage}[t]{0.32\textwidth}
\centering
\includegraphics[width=1\linewidth]{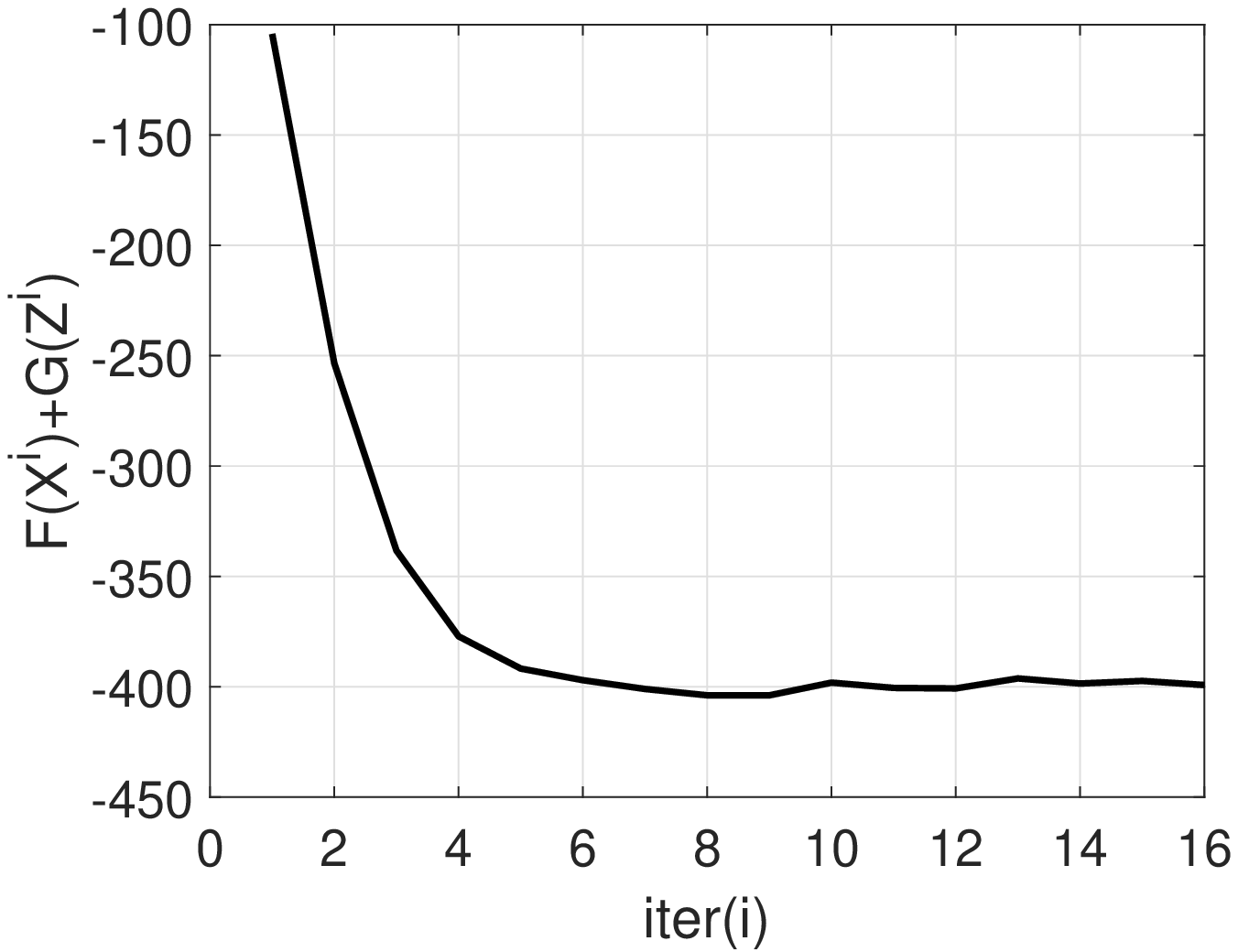}
\caption{Convergence of objective $\rm{{ }}$ value.}
\label{fig_fxgx}
\end{minipage}%
\begin{minipage}[t]{0.32\textwidth}
\centering
\includegraphics[width=1\linewidth]{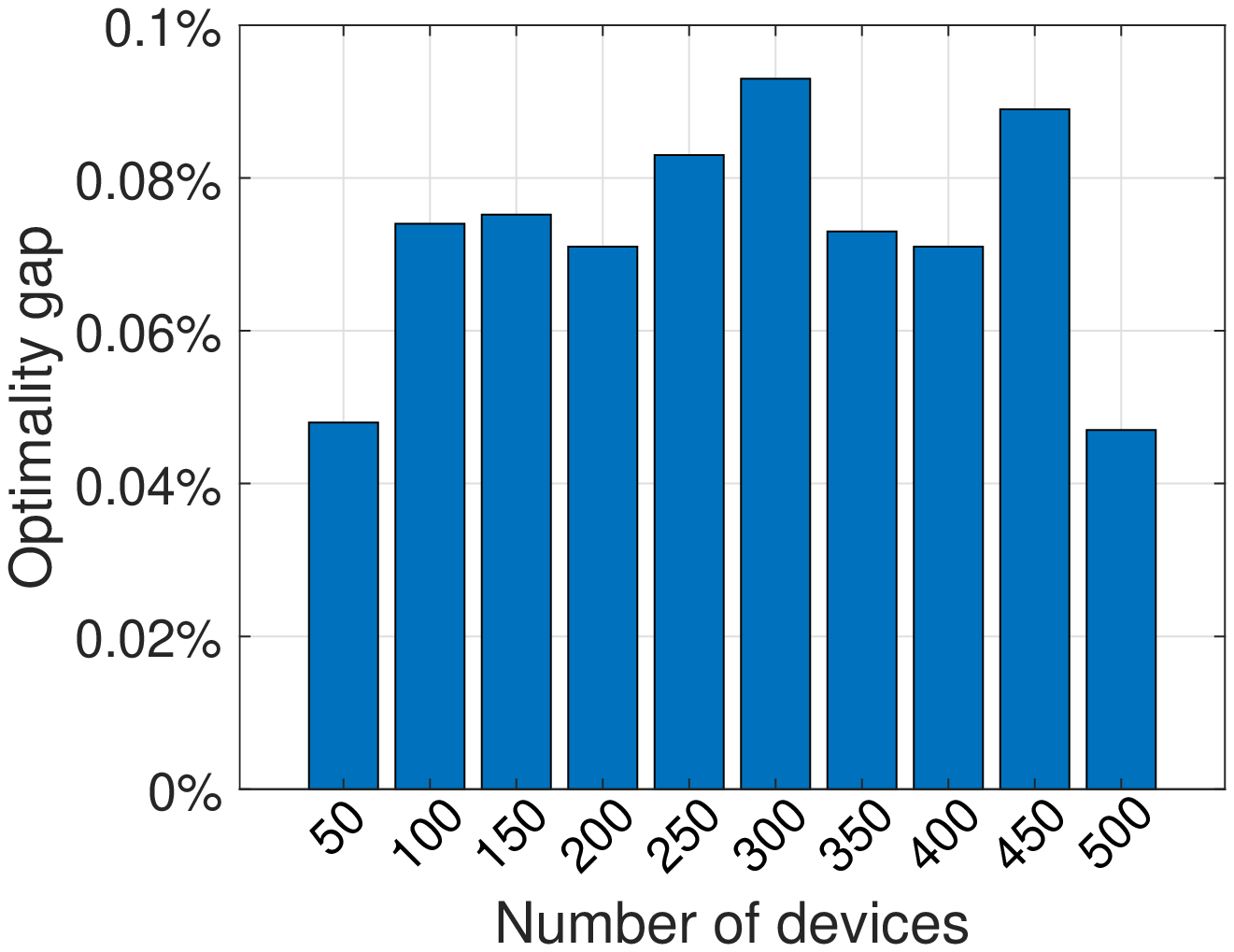}
\caption{Optimality gap versus the number of devices.}
\label{fig_ADMMobj}
\end{minipage}
\begin{minipage}[t]{0.32\textwidth}
\centering
\includegraphics[width=1\linewidth]{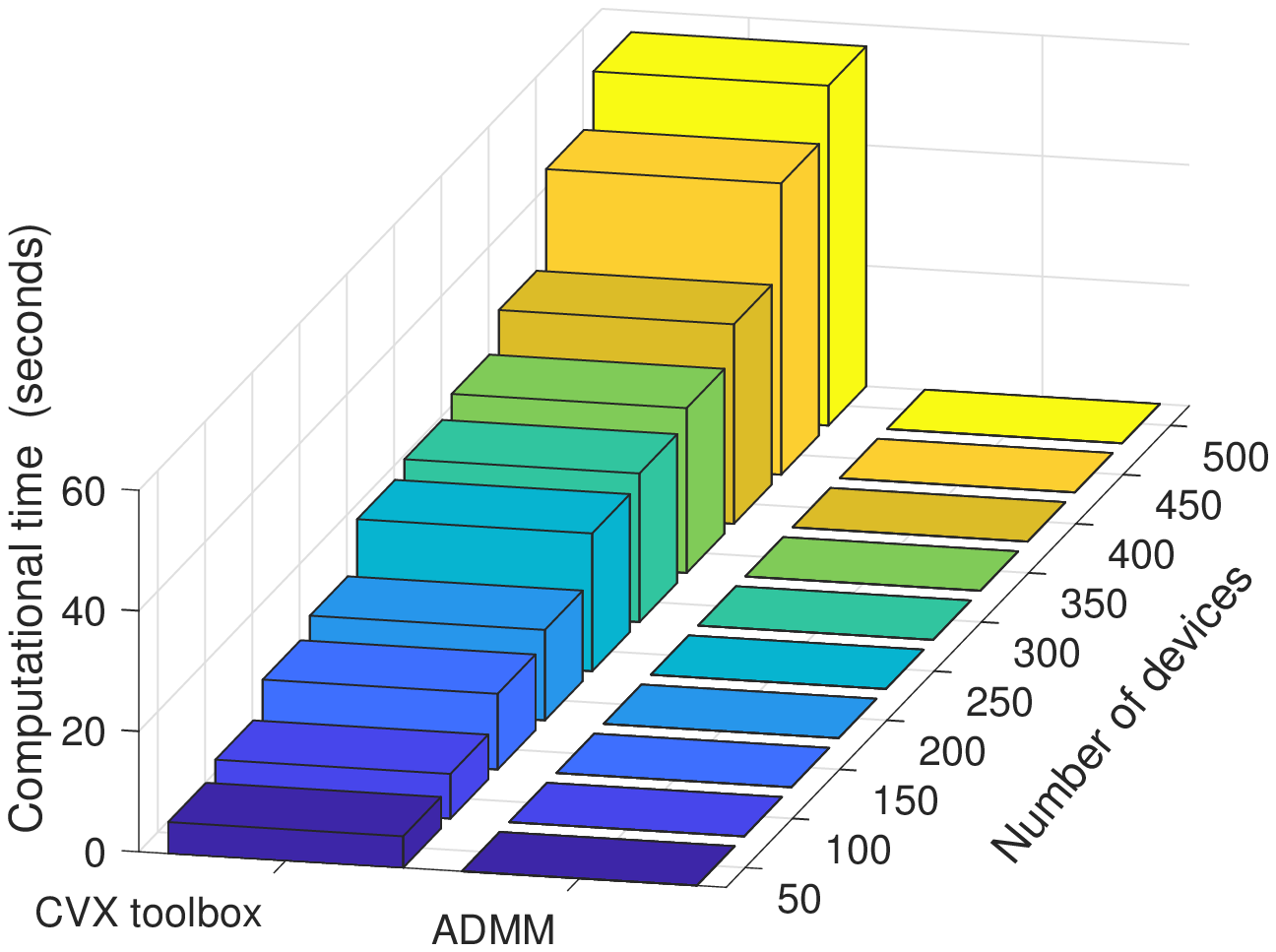}
\caption{Computational time versus the number of devices.}
\label{fig_ADMMtime}
\end{minipage}
\end{figure*}

\subsection{Data Queue Performance}
\label{sec:71}

\subsubsection{Data queue backlog}
Fig. \ref{fig3:side:3} and Fig. \ref{fig3:side:4} show the evolutions of data queue backlog corresponding to the proposed algorithm and the baseline $1$ algorithm, respectively. The data queue backlog of the proposed algorithm tends to be stable within a short period of time, which guarantees reliable service provision, while the data queue backlog of the baseline $1$ algorithm fluctuates more violently, thus making the network less stable. Compared with the baseline $1$ algorithm, the proposed algorithm can reduce the peak to average ratio (PAR) of data queue backlog by $35.5 \%$. This phenomenon has also been validated in Fig. \ref{fig3:side:5}, which shows the empirical cumulative distribution function (CDF) performance of data queue backlog. Taking device $1$ as an example, the probability that the data queue backlog $Q_1$ lies within the region $[2.501, 2.548]$ is $0.5604$, while the probability corresponding to the baseline $1$ algorithm is only $0.1608$.

\subsubsection{Data arrival rate}

The data arrival rate performances of the proposed algorithm and the baseline $1$ algorithm are shown in Fig. \ref{fig4:side:6} to Fig. \ref{fig4:side:8}. Similar to the data queue backlog performance, the data arrival rate fluctuation of the proposed algorithm is much less than that of the baseline $1$ algorithm. The proposed algorithm can reduce the PAR of data arrival rate by $28.5 \%$, which infers a more stable rate control performance.

\subsubsection{Data transmission rate}
Fig. \ref{fig5:side:11} shows the shaded error bar of data transmission rate, where the width of the shadow represents the standard deviation. Compared with the baseline $1$ algorithm, the proposed algorithm has a much narrower shadow, which provides a more stable transmission rate. Furthermore, we can find that the proposed algorithm can differentiate devices by providing higher transmission rate to devices with larger service weights. In comparison, the baseline $1$ algorithm treats all the devices as if they have the same service weight.

\subsubsection{Queuing delay}
Fig. \ref{fig_delay_BAR} shows the time-average queuing delay performance. For the proposed algorithm, we can find that devices with higher service priorities, e.g., device $5$, experience less delay compared with devices with lower service priorities, e.g., device $1$. This is consistent with the transmission rate results shown in Fig. \ref{fig5:side:11}. For example, since it has the highest priority, the time-average data output of device $5$ is also the highest among those of all the devices in order to satisfy (5). Considering the fact that the queuing delay is inversely proportional to the time-average data output, device $5$ may have the smallest queuing delay based on (5) and (6). While for the baseline $1$ algorithm, delay is uncorrelated with service priority, thus making differentiated service provisioning impossible. 

\subsubsection{Network utility}
Fig. \ref{fig11:side:24} and Fig. \ref{fig11:side:25} show the network utility performances of the proposed algorithm and the baseline $1$ algorithm, respectively. The proposed algorithm achieves more stable network utility performance than the baseline $1$ algorithm, which implies a better adaptiveness to the random variation of channel state.

\subsection{Energy Queue Performance}
\label{sec:72}

\subsubsection{Energy consumption}

 {Fig. \ref{fig_BETA5000_adapt}-Fig. \ref{twoprice2} show the relations between electricity price and purchased grid energy for the proposed algorithm and the baseline $2$ algorithm respectively. In Fig. \ref{fig_BETA5000_adapt} and Fig. \ref{fig_BETA0_adapt}, we adopt a sinusoid-based electricity price with the minimum value of $1.8$ RMB/kWh and the maximum value of $9.0$ RMB/kWh. The choice of sinusoid function has two advantages: 1) it highlights the fluctuation of the electricity price; 2) it characterizes the peak and valley features of the electricity price. Similar electricity price model is also adopted in \cite{ZhangDistributed}. In Fig. \ref{twoprice1} and Fig. \ref{twoprice2}, we adopt a two-tier electricity price with the minimum value of $3$ RMB/kWh and the maximum value of $7$ RMB/kWh. Simulation results demonstrate that the proposed algorithm can dynamically adapt purchased energy with time-varying electricity price by avoiding purchasing the expensive grid power during the peak-price period. In comparison, the baseline $2$ algorithm is unaware of the energy cost, and consumes grid energy any time if needed. This inevitably leads to higher energy cost, which is demonstrated in Fig. \ref{fig_Energycost} and Fig. \ref{fig_Energycost_BAR}.}

{Fig. \ref{kappa_price} shows the total purchased energy and the total harvested energy over $200$ energy frames versus different values of $\kappa$. The simulation results demonstrate that the optimal values of ${\vartheta }\left[ {(m - 1)T + 1} \right]$ and ${g}\left[ {(m - 1)T + 1} \right]$ do not depend on ${\kappa }\left[ {(m - 1)T + 1} \right]$. }

\subsubsection{Grid energy cost}

Fig. \ref{fig_Energycost} and Fig. \ref{fig_Energycost_BAR} show the grid energy cost per energy frame and the total energy cost accumulated over $200$ energy frames, respectively. Compared with the baseline $2$ algorithm, the proposed algorithm can reduce the energy cost by $48.23\%$, due to the awareness of electricity price and dynamic adaptation of gird energy consumption.

\subsubsection{QoE and energy backlog}

Fig. \ref{fig_QOE_BAR} and Fig. \ref{fig_EQEQ_BAR} show the network QoE and energy queue backlog, respectively, where the bar graph represents the time-average value and the error bar represents standard deviation. It is observed that time-average QoE performance of the proposed algorithm is only $0.54\%$ lower than the that of baseline $2$ algorithm, while the energy queue backlog is only $5.40\%$ lower. In other words, the proposed algorithm trades only $0.54\%$ QoE performance degradation and $5.40\%$ energy queue backlog reduction for energy cost reduction as high as $48.23\%$.

\subsection{Computational Complexity and Convergence Performances}
\subsubsection{Convergence of Algorithm \ref{ADMMoptimization}}
Fig. \ref{fig_admmRP}, Fig. \ref{fig_admmRD}, and Fig. \ref{fig_fxgx} show the primal residual convergence, dual residual convergence, and optimal convergence of the ADMM-based rate control algorithm at $\tau=1000$, respectively. It is observed that the stopping criterion constraints $\epsilon^{pri}$ and $\epsilon^{dual}$, i.e., the dotted lines shown in Fig. \ref{fig_admmRP} and Fig. \ref{fig_admmRD}, can be satisfied within $16$ iterations. Fig. \ref{fig_fxgx} demonstrates that the objective value converges to the optimal value within only 7 iterations when there are 500 devices.
\subsubsection{Optimality gap and computational complexity} 
Fig. \ref{fig_ADMMobj} and Fig. \ref{fig_ADMMtime} compare optimality and computational complexity between the proposed algorithm and the baseline $3$ algorithm, respectively. Simulation results demonstrate that the optimality gap between the two algorithms is always less than $1 \%$. On the other hand, the proposed algorithm can reduce the computational time by $99 \%$ compared with the baseline $3$ algorithm when there are $500$ devices. Furthermore, the computational time of the baseline $3$ algorithm increases significantly with the number of devices, while that of the proposed algorithm remains in a much lower level.

\section{Conclusions}
\label{sec:8}

In this paper, we studied the two-timescale resource allocation problem in 5G-empowered automated networks for IIoT applications. We proposed a two-timescale resource allocation algorithm to maximize the long-term QoE performance while simultaneously minimizing the grid energy cost, in which the optimization of energy management is performed every energy frame, while the optimization of rate control, channel selection, and power allocation is performed every data slot. We proved that the proposed algorithm can achieve bounded performance deviation based only on causal information of CSI, EH, and electricity price. We compared it with three heuristic algorithms under various simulation configurations. Simulation results demonstrate that the proposed algorithm can effectively reduce the PARs of data queue backlog and data arrival rate by $35.5\%$ and $28.5 \%$, respectively. It allows differentiated service provision and achieves $48.23\%$ energy cost reduction by dynamically adapting resource allocation with service priority and time-varying electricity price. It is able to trade only $1 \%$ optimality performance degradation for $99 \%$ computatational time reduction. In the future work, we plan to study how to adopt the machine learning with existing framework to further improve the performance.
%%%%%%%%%%%%%%%%%%%%%%%%%%%%%%%%%%%%%%%%%%%%%%%%%%%%%%%%%%%%%%%%%%%%%%%%%%%%%%%%%%%%%%%%%%%%%%%%%%%%%%%%%%%%%%%%%%%%%%%%%%%%%%%%%%%%%%
%%%%%%%%%%%%%%%%%%%%%%%%%%%%%%%%%%%%%%%%%%%%%%%%%%%%%%%%%%%%%%%%%%%%%%%%%%%%%%%%%%%%%%%%%%%%%%%%%%%%%%%%%%%%%%
\appendices
\section{Proof of Theorem 1}
\label{APPA}
For any nonnegative real numbers $Q_n\left( {\tau} \right)$, $r_n\left( {\tau} \right)$ and $v_n\left( {\tau} \right)$, there holds
\begin{align}\label{eq32}
&\dfrac{1}{2}\left[ {Q_n^2\left( {\tau + 1} \right) - Q_n^2\left( \tau \right)} \right] \nonumber \\
\le& \dfrac{1}{2}r_n^2\left( \tau \right){T_0^2} + \dfrac{1}{2}v_n^2\left( \tau \right){T_0^2}
+{Q_n}\left( \tau \right)\left[ {{r_n}\left( \tau \right) - {v_n}\left( \tau \right)} \right]{T_0}.
\end{align}

Applying the law of telescoping sums over $\tau \in \left[{(m-1)T+1,mT}\right]$, we can derive
\begin{align}\label{eq34}
&\dfrac{1}{2}\left[ {Q_n^2\left( {\tau + T} \right) - Q_n^2\left( \tau \right)} \right] \nonumber \\
\le &\dfrac{1}{2}\displaystyle\sum\limits_{\tau = (m-1)T+1}^{mT} {\left[r_n^2\left( \tau \right)+ v_n^2\left( \tau \right)\right]{T_0^2} } \notag\\
+&\displaystyle\sum\limits_{\tau = (m-1)T+1}^{mT} {\left\{{Q_n}\left( \tau \right)\left[ {{r_n}\left( \tau \right) - {v_n}\left( \tau \right)} \right]\right\}T_0}.
\end{align}

Similarly, for energy queue, we can derive
\begin{align}\label{eq35}
&\dfrac{1}{2}\left[ {{{\cal E}^2}\left( {\tau{\rm{ + }}1} \right) - {{\cal E}^2}\left( \tau \right)} \right] \nonumber \\
\le& \dfrac{1}{2}{p_{_{c}}^2\left( \tau \right)+\left[ {{g}\left( \tau \right) + \vartheta \left( \tau \right)} \right]^2} + {\cal E}\left( \tau \right)\left\{ {{p_{c}}\left( \tau \right) - \left[ {{g}\left( \tau \right) + \vartheta \left( \tau \right)} \right]} \right\}.
\end{align}

\begin{align}\label{eqnew}
&\dfrac{1}{2}\left[ {{{\cal E}^2}\left( {\tau{\rm{ + }}T} \right) - {{\cal E}^2}\left( \tau \right)} \right] \nonumber \\
\le &
\dfrac{1}{2}\sum\limits_{\tau = (m-1)T+1}^{mT}\left[{p_{_{c}}^2\left( \tau \right)+\left[ {{g}\left( \tau \right) + \vartheta \left( \tau \right)} \right]^2}\right] \notag \\
+&\dfrac{1}{2}\sum\limits_{\tau = (m-1)T+1}^{mT} {\cal E}\left( \tau \right)\left\{ {{p_{c}}\left( \tau \right) - \left[ {{g}\left( \tau \right) + \vartheta \left( \tau \right)} \right]} \right\}.
\end{align}

Combining (\ref{eq34}) and (\ref{eqnew}) as well as applying the law of telescoping sums and the law of iterated expectations, we derive

\begin{align}
\label{eqbound1}
&{\Delta_T}\left( \tau \right)\le\dfrac{1}{2}BT \nonumber \\
+& \displaystyle\sum\limits_{\tau = (m-1)T+1}^{mT} {\displaystyle\sum\limits_{n = 1}^N {\mathbb{E}\left\{ {{T_0}{Q_n}\left( \tau \right)\left[ {{r_n}\left( \tau \right) - {v_n}\left( \tau \right)} \right]\left| {H\left( \tau \right)} \right.} \right\}} }\notag\\
+& \displaystyle\sum\limits_{\tau = {(m - 1)T+1}}^{mT}{\mathbb{E}\left\{ {{\cal{E}}\left( \tau \right)\left\{ {{p_c}\left( \tau \right) - \left[ {g\left( \tau \right) + \vartheta \left( \tau \right)} \right]} \right\}\left| {H\left( \tau \right)} \right.} \right\}}.
\end{align}
where ${\Delta_T}\left( \tau \right)=\dfrac{1}{2}\left[ {Q_n^2\left( {\tau + T} \right) - Q_n^2\left( \tau \right)} \right]+\dfrac{1}{2}\left[ {{{\cal E}^2}\left( {\tau{\rm{ + }}T} \right) - {{\cal E}^2}\left( \tau \right)} \right]$. Based on the definition of DMU, we can subtract the term ${\mathbb{E}\left[ Vf\left( \tau \right)\left| {H\left( \tau \right)} \right.\right]}$ from both sides of (\ref{eqbound1}), and then apply the law of iterated expectations to derive the upper bound of DMU, which is given by
\begin{align}
D[ {{\cal H}( \tau )} ]\le& \dfrac{1}{2}BT 
+\displaystyle\sum\limits_{\tau = (m-1)T+1}^{mT}{\mathbb{E} \{\{  V\beta \eta ( t )[{g}( t )+\vartheta(t)]}\notag\\
-&{\cal{E}(\tau)}[ {{g}( \tau ) + \vartheta ( \tau )} ]\}|{{\cal H}( \tau )} .\} \notag\\
+&\sum\limits_{\tau = (m-1)T+1}^{mT} \sum\limits_{n = 1}^N \mathbb{E} \{\{ [ Q_n(\tau ){r_n}( \tau ){T_0} ] \notag\\
-& V{U_n}( \tau ) \}|\cal H( \tau ) .\} \notag\\
-&\displaystyle\sum\limits_{\tau = (m-1)T+1}^{mT} \displaystyle\sum\limits_{n = 1}^N\sum\limits_{k = 1}^K x_{n,k} \mathbb{E}\{\{ {Q_n}( \tau ){v_n}( \tau ){T_0} \notag\\
-&{\cal{E}}( \tau ) {p_{n,k}}( \tau ) \}|{\cal H}( \tau ) \} \label{eq38}.
\end{align}

Define ${D_0}\left( \tau \right)$ as
\begin{align}
{D_0}( \tau )=&\sum\limits_{\tau = (m-1)T+1}^{mT}\mathbb{E} \{\{  V\beta \eta ( t )[g( t )+\vartheta(t)]{p_{n,k}} \notag\\
-&\cal{E}(\tau)[ {g}( \tau ) + \vartheta ( \tau ) ]\}|{{\cal H}( \tau )} .\}.
\end{align}

According to the energy causality, for $\forall{\tau^{'}} > {\tau}$, the following inequality holds
\begin{align}\label{eqEEE}
&E\left( {{\tau}} \right) - \left( {{\tau^{'}} - {\tau}} \right){p_{c }} \le E\left( {{\tau^{'}}} \right) \notag\\\le& E\left( {{\tau}} \right)+\left( {{\tau^{'}} - {\tau}} \right)\left( {{{g}_{\max }} + {\vartheta _{\max }}} \right),
\end{align}
where $\vartheta_{\max}$ is the upper bound of ${\mathbb{E}}\left[ {\vartheta\left( \tau \right)\left| {{\cal H}\left( \tau \right)} \right.} \right]$.

Using these inequalities (\ref{eqEEE}) over $\tau \in \left[{(m-1)T+1,mT}\right]$, we can derive
\begin{align}\label{eqD0}
{D_0}\left( \tau \right)
\le& \dfrac{{\left( {T - 1} \right)T}}{2}{\left( {{{g}_{\max }}+ {\vartheta _{\max }}} \right)^2} \notag\\
+& TV\beta \eta \left[ {(m-1)T+1} \right]\{{g}\left[ {(m-1)T+1} \right]\notag\\
+&{\vartheta}\left[ {(m-1)T+1} \right] \}\notag\\
+& {\cal{E}}\left[ {(m-1)T+1} \right]\{ {g}\left[ {(m-1)T+1} \right] \notag\\
+& \vartheta \left[ {(m-1)T+1} \right] \}.
\end{align}

This completes the proof of Theorem 1.

\section{Proof of Theorem 2}
\label{APPB}

To prove Theorem 2, we introduce some significant and practical assumptions, i.e.,
\begin{align}
\mathbb{E}\left[ {{r_n}\left( \tau \right){T_0} - {v_n}\left( \tau \right){T_0}}{\left| {{Q_n}\left( \tau \right)} \right.} \right] \le& - {\delta _1},\\
\mathbb{E}\left\{ {{p_c}\left( \tau \right) - \left[ {g\left( \tau \right) + \vartheta \left( \tau \right)} \right]}{\left| {{E}\left( \tau \right)} \right.} \right\} \le& - {\delta _2},
\end{align}
where ${\delta _1}>0$ and ${\delta _2}>0$ are the gap between data queue input and output, and the gap between energy queue input and output, respectively.

According to Theorem 1, we can derive
\begin{align}
D\left[ {{\cal H}\left( \tau \right)} \right]\le& \displaystyle\frac{1}{2}BT- V{f_{opt}}\notag\\
+ &\displaystyle\sum\limits_{\tau = (m-1)T+1}^{mT} {\displaystyle\sum\limits_{n = 1}^N{\mathbb{E}\left\{{Q_n}\left( \tau \right)\left[ {{r_n}\left( \tau \right) - {v_n}\left( \tau \right)} \right]\right\}} }\notag\\
+ &\displaystyle\sum\limits_{\tau = (m-1)T+1}^{mT}{\mathbb{E}\left\{{\cal E}\left( \tau \right)\left\{ {{p_{c}}\left( \tau \right) - \left[ {{g}\left( \tau \right) + \vartheta \left( \tau \right)} \right]} \right\}\right\}}.
\end{align}

Applying the law of telescoping sums over $\tau \in \left[{1,MT}\right]$ and the law of iterated expectations for the above equation, we derive
\begin{align}
&\mathbb{E}\left[{L\left( {MT} \right) - L\left( 1 \right) - V\sum\limits_{m = 1}^M {\sum\limits_{\tau = (m - 1)T + 1}^{mT} {f\left( \tau \right)} } }\right] \le \dfrac{1}{2}MBT \notag\\
- &\mathbb{E} \left[{\sum\limits_{m = 1}^M {\sum\limits_{\tau = \left( {m - 1} \right)T + 1}^{mT} {\sum\limits_{n = 1}^N {{Q_n}\left( \tau \right){\delta _1}} } } }\right] \notag\\
-& \mathbb{E} \displaystyle\sum\limits_{m = 1}^M {\sum\limits_{\tau = \left( {m - 1} \right)T + 1}^{mT} {E\left( \tau \right){\delta _2}} } - VMT{f_{opt}} \label{eq45}.
\end{align}

According to (\ref{eq45}), we can derive
\begin{align}
&\mathbb{E}\left[ {L\left( {MT} \right) - L\left( 1 \right) - V\displaystyle\sum\limits_{m = 1}^M \displaystyle\sum\limits_{\tau = \left( {m - 1} \right)T + 1}^{mT}{f\left( \tau \right)} } \right]\notag\\
\le& \displaystyle \frac{1}{2}MBT - \mathbb{E}\displaystyle \sum\limits_{m = 1}^M {\sum\limits_{\tau = \left( {m - 1} \right)T + 1}^{mT} {\sum\limits_{n = 1}^N {{Q_n}\left( \tau \right){\delta _1}} } } - VMT{f_{opt}} \label{eq46}.
\end{align}

Rearranging (\ref{eq46}), we obtain
\begin{align}
&\mathbb{E} \left[ \displaystyle \sum\limits_{m = 1}^M {\sum\limits_{\tau = \left( {m - 1} \right)T + 1}^{mT} {\sum\limits_{n = 1}^N {{Q_n}\left( \tau \right){\delta _1}} } } \right]\notag\\
\le&\displaystyle \frac{1}{2}MBT - \mathbb{E}\left[ {L\left( {MT} \right) - L\left( 1 \right)} \right] + VMT\left( {{f_{\max }} - {f_{opt}}} \right)\label{eq47},
\end{align}
{{where ${f_{\max }}$ is the finite constant to bound $\mathbb{E}\left[f(\tau)\right]$, and $f_{opt}$ is the theoretical optimum of \bf{P1}.}} There exists a bound that ${f_{\max }} \ge {f_{opt}}$.

Dividing both sides of (\ref{eq47}) by $MT{\delta _1}$ and taking the limit $M \to \infty$, we can obtain
\begin{align}
&\mathop {\lim }\limits_{M \to \infty } \displaystyle\frac{1}{{MT}}\mathbb{E} \left[\displaystyle\sum\limits_{m = 1}^M {\sum\limits_{\tau = \left( {m - 1} \right)T + 1}^{mT} {\sum\limits_{n = 1}^N {{Q_n}\left( \tau \right)} } }\right] \notag\\
\le& \frac{1}{{2{\delta _1}}}B + \frac{{V\left( {{f_{\max }} - {f_{opt}}} \right)}}{{{\delta _1}}}.
\end{align}

Similarly, based on (\ref{eq45}), we can obtain
\begin{align}\label{eq51}
&\displaystyle \frac{1}{{MT}}\mathbb{E}\left[\displaystyle \sum\limits_{m = 1}^M {\displaystyle \sum\limits_{\tau = \left( {m - 1} \right)T + 1}^{mT} {{\cal E}\left( \tau \right)} } \right] \notag\\
\le& \displaystyle \frac{1}{{2{\delta _2}}}B - \displaystyle \frac{{\mathbb{E}\left[ {L\left( {MT} \right) - L\left( 1 \right)} \right]}}{{MT{\delta _2}}} + \displaystyle \frac{{V\left( {{f_{\max }} - {f_{opt}}} \right)}}{{{\delta _2}}}.
\end{align}

Taking the limit $M \to \infty$ and using $\mathop {\lim }\limits_{M \to \infty } \displaystyle\frac{{\mathbb{E}\left[ {L\left( {MT} \right) - L\left( 1\right)} \right]}}{{MT{\delta _2}}} = 0$, we can derive
\begin{align}\label{eq52}
\mathop{\lim }\limits_{M\to\infty }\frac{1}{{MT}}\mathbb{E}\displaystyle\sum\limits_{m = 1}^M {\displaystyle\sum\limits_{\tau = \left( {m - 1} \right)T + 1}^{mT} {{\cal E}\left( \tau\right)}} \le\displaystyle\frac{1}{{2{\delta _2}}}B+\frac{{V\left({{f_{\max }}-{f_{opt}}}\right)}}{{{\delta _2}}}.
\end{align}

Rearranging (\ref{eq52}), we have
\begin{align}
&\mathop{\lim }\limits_{M\to\infty }\displaystyle\frac{1}{{MT}}\mathbb{E}\displaystyle\sum\limits_{m = 1}^M {\displaystyle\sum\limits_{\tau = \left( {m - 1} \right)T + 1}^{mT} {{E}\left( \tau\right)}} \ge {E_{\max}}\notag\\
-&\displaystyle\frac{1}{{2{\delta _2}}}B-\frac{{V\left({{f_{\max }}-{f_{opt}}}\right)}}{{{\delta _2}}}.
\end{align}

Similarly, based on (\ref{eq45}), we can obtain
\begin{align}\label{eq53}
&\mathop {\lim }\limits_{M \to \infty } \dfrac{1}{{MT}}\mathbb{E}\left[ {\sum\limits_{m = 1}^M {\sum\limits_{\tau = \left( {m - 1} \right)T{\rm{ + }}1}^{mT} {f\left( \tau \right)} } } \right] \notag\\
\ge& \mathop {\lim }\limits_{M \to \infty } \dfrac{1}{{VMT}}\mathbb{E}\left[ {L\left( {MT} \right) - L\left( 1 \right)} \right] - \dfrac{B}{{2V}} + {f_{opt}}.
\end{align}

Since $ \mathop {\lim }\limits_{M \to \infty }\dfrac{1}{{VMT}}\mathbb{E}\left[ {L\left( {MT} \right) - L\left( 1 \right)} \right]=0$, we have
\begin{align}
\mathop {\lim }\limits_{M \to \infty } \dfrac{1}{{MT}}\mathbb{E}\left[ {\sum\limits_{m = 1}^M {f\left( \tau \right)} } \right] \ge {f_{opt}} -\dfrac{B}{{2V}}.
\end{align}

This completes the proof of Theorem 2.

\section{Proof of Theorem 3}
\label{APPC}
The contradiction method is utilized to prove Theorem $3$. Assume that there exists a blocking pair $\left(n, k \right)$ in the final matching $\varphi^*$, which means that $\varphi^*(n)=k$ but $n$ and $k$ would prefer to disrupt the matching in order to be matched with each other. According to \textbf{Definition 2}, the matching does not terminate until all the blocking pairs are eliminated. In other words, $\varphi^*$ is not the final matching, which causes conflict with the assumption. Therefore, there does not exist a blocking pair in the final matching, and the proposed algorithm produces a stable matching between devices and channels within finite iterations.

\section{Proof of Theorem 4}
\label{APPD}
The properties of residual convergence, objective convergence, and dual variable convergence specified in \textbf{Theorem $\mathbf{4}$} hold if the objective function of $\mathbf{P7}$, i.e., $F_r(\mathbf{x}_r)+G_r(\mathbf{z}_r)$, is closed, proper, and convex, and the Lagrangian $ {\bf{L}}_{\rho}({\bf{x}}_r, {\bf{y}}_r, {{y}})$ has a saddle point.

$F_r(\mathbf{x}_r)$ is composed of two convex functions. According to the accumulation nature of convex functions, $F_r(\mathbf{x}_r)$ is convex. Similarly, based on (31), $G_r(\mathbf{z}_r)$ is also convex. 
Therefore, $F_r(\mathbf{x}_r)+G_r(\mathbf{z}_r)$ is convex. 

Second, since $F_r(\mathbf{x}_r)$ is convex, there exists at least one minimum point. Therefore, the set of $\left\{r_{l_r}(\tau)|r_{l_r}(\tau)\in {\rm dom}\ F_r(\mathbf{x}_r);F_r(\mathbf{x}_r) \leq \alpha \right\}$, $\forall \alpha \in \mathbb{R}$, is closed, i.e., $ F_r(\mathbf{x}_r)$ is closed. Similarly, $G_r(\mathbf{z}_r)$ is also closed. Thus, $F_r(\mathbf{x}_r)+G_r(\mathbf{z}_r)$ is closed.

Third, there must exist at least one set of variables $\mathbf{x}_r \in \mathbb{R}^{l_r\times 1}$ which satisfy $F_r(\mathbf{x}_r)<\infty$. On the other hand, due to the existence of the minimum point, $F_r(\mathbf{x}_r)>-\infty$ is true for all $\mathbf{x}_r \in \mathbb{R}^{l_r\times 1}$. That is, $F_r(\mathbf{x}_r)$ is proper. Similarly, we can prove that $G_r(\mathbf{z}_r)$ is also proper. Therefore, $F_r(\mathbf{x}_r)+G_r(\mathbf{z}_r)$ is proper.

Define $(\mathbf{x}_r^*,\mathbf{z}_r^*)$ as the optimal value of the primal problem, which is given by 
\begin{align}
\left(\mathbf{x}_r^*,\mathbf{z}_r^* \right)=\inf\limits_{(\mathbf{x}_r,\mathbf{z}_r)} \sup \limits_{y} \mathbf{L}_\rho \left(\mathbf{x}_r,\mathbf{z}_r,y\right).
\end{align}

Define $y^*$ as the optimal value of the dual problem, which is given by
\begin{align}
y^*= \sup \limits_{y}\inf\limits_{(\mathbf{x}_r,\mathbf{z}_r)} \mathbf{L}_\rho \left(\mathbf{x}_r,\mathbf{z}_r,y\right).
\end{align}

Since $(\mathbf{x}_r^*,\mathbf{z}_r^*)$ minimizes $\mathbf{L}_\rho \left(\mathbf{x}_r,\mathbf{z}_r,y^*\right)$ over ${{\bf{x}}_r} \in {{\mathbb{R}}^{{l_r}\times 1}}$ and ${\bf{z}}_r \in {{\mathbb{R}}^{\left( {N- {l_r}} \right)\times 1}}$, and $y^*$ maximizes $\mathbf{L}_\rho \left(\mathbf{x}_r^*,\mathbf{z}_r^*,y\right)$ over $y \leq 0$, we have 
\begin{align}
\mathbf{L}_\rho \left(\mathbf{x}_r^*,\mathbf{z}_r^*,y^*\right)&=\inf \limits_{(\mathbf{x}_r,\mathbf{z}_r)} \mathbf{L}_\rho \left(\mathbf{x}_r,\mathbf{z}_r,y^*\right), \\
\mathbf{L}_\rho \left(\mathbf{x}_r^*,\mathbf{z}_r^*,y^*\right)&=\sup \limits_{y} \mathbf{L}_\rho \left(\mathbf{x}_r^*,\mathbf{z}_r^*,y\right).
\end{align}

Then, we have 
\begin{align}
\label{SPADMM}
\mathbf{L}_\rho \left(\mathbf{x}_r^*,\mathbf{z}_r^*,y\right)\leq \mathbf{L}_\rho \left(\mathbf{x}_r^*,\mathbf{z}_r^*,y^*\right)\leq \mathbf{L}_\rho \left(\mathbf{x}_r,\mathbf{z}_r,y^*\right), 
\end{align}

Based on (\ref{SPADMM}), $\left(\mathbf{x}_r^*,\mathbf{z}_r^*,y^*\right)$ is a saddle point for $\mathbf{L}_\rho \left(\mathbf{x}_r,\mathbf{z}_r,y\right)$. 

Therefore, the properties of residual convergence, objective convergence, and dual variable convergence hold because the objective function of $\mathbf{P7}$ is closed, proper, and convex, and the Lagrangian $ {\bf{L}}_{\rho}({\bf{x}}_r, {\bf{y}}_r, {{y}})$ has a saddle point. 

\bibliographystyle{IEEEtran}% mathematics and physical sciences
\bibliography{hyhno7}

\end{document}